\documentclass[twocolumn]{aastex63}

\newcommand{\source}{SGR 1935+2154}

\shorttitle{A Galactic FRB in the Context of Magnetar Models}
\shortauthors{Margalit, Beniamini, Sridhar, \& Metzger}
\graphicspath{{./}{figures/}}

\begin{document}

\title{\Large{Implications of a ``Fast Radio Burst" from a Galactic Magnetar}}

\correspondingauthor{Ben Margalit}
\email{benmargalit@berkeley.edu}

\author{Ben Margalit}
\altaffiliation{NASA Einstein Fellow}
\affiliation{Astronomy Department and Theoretical Astrophysics Center, University of California, Berkeley, Berkeley, CA 94720, USA}

\author{Paz Beniamini}
\affiliation{Division of Physics, Mathematics and Astronomy, California Institute of Technology, Pasadena, CA 91125, USA}

\author{Navin Sridhar}
\affiliation{Department of Astronomy, Columbia University, New York, NY 10027, USA}

\author{Brian D. Metzger}
\affiliation{Columbia Astrophysics Laboratory, Columbia University, New York, NY 10027, USA}
\affiliation{Center for Computational Astrophysics, Flatiron Institute, 162 W. 5th Avenue, New York, NY 10011, USA}


\begin{abstract}

A luminous radio burst was recently detected in temporal coincidence with a hard X-ray flare from the Galactic magnetar \source~with a time and frequency structure consistent with cosmological fast radio bursts (FRB) and a fluence within a factor of $\lesssim 10$ of the least energetic extragalactic FRB previously detected.  Although active magnetars are commonly invoked FRB sources, several distinct mechanisms have been proposed for generating the radio emission which make different predictions for the accompanying higher frequency radiation.  We show that the properties of the coincident radio and X-ray flares from \source, including their approximate simultaneity and relative fluence $E_{\rm radio}/E_{\rm X} \sim 10^{-5}$, as well as the duration and spectrum of the X-ray emission, are consistent with extant predictions for the synchrotron maser shock model.  Rather than arising from the inner magnetosphere, the X-rays are generated by (incoherent) synchrotron radiation from thermal electrons heated at the same internal shocks which produce the coherent maser emission as ultra-relativistic flare ejecta collides with a slower particle outflow (e.g. as generated by earlier flaring activity) on a radial scale $\sim 10^{11}$ cm.  Although the rate of \source-like bursts in the local universe is not sufficient to contribute appreciably to the extragalactic FRB rate, the inclusion of an additional population of more active magnetars with stronger magnetic fields than the Galactic population can explain both the FRB rate as well as the repeating fraction,
however only if the population of active magnetars are born at a rate that is at least two-orders of magnitude lower than that of \source-like magnetars.  This may imply that the more active magnetar sources are not younger magnetars formed in a similar way to the Milky Way population (e.g. via ordinary supernovae), but instead through more exotic channels such as superluminous supernovae, accretion-induced collapse or neutron star mergers.
\end{abstract}

\keywords{
Radio transient sources (2008) --- Magnetars (992) --- Soft gamma-ray repeaters (1471)
}


\section{Introduction} 
\label{sec:intro}

Fast radio bursts (FRBs) are bright, millisecond duration pulses of coherent radio emission with dispersion measures (DM) well in excess of Galactic values (\citealt{Lorimer+07,Thornton+13}; see \citealt{Petroff+19}, \citealt{Cordes&Chatterjee19} for reviews), and hence pointing to an extragalactic origin. The precise mechanisms powering FRBs remain a topic of debate, in large part due to the small number of well localized events, as well as the fact that some FRBs appear to repeat \citep{Spitler+16,Chatterjee+17,CHIME_repeaters} while others do not. 

Many theoretical models have been proposed for FRBs (see \citealt{Platts+19} for a catalog).  Perhaps the most well-studied models are those which postulate that FRBs arise from the flaring activity of strongly-magnetized neutron stars (NS) known as ``magnetars" \citep{Popov&Postnov13,Lyubarsky14,Kulkarni+14,Katz16,Metzger+17,Beloborodov17,Kumar+17}.  Evidence in favor of magnetars as FRB sources include: (1) high linear polarization and large rotation measures (e.g.~\citealt{Masui+15,Michilli+18}), indicative of a strongly-magnetized central engine and environment; (2) spatial association with star-forming regions, in the two repeating events where VLBI imaging enables precise sky localizations \citep{Bassa+17,Tendulkar+17,Marcote+20}; (3) statistical properties of the bursts' repetition consistent with those of Galactic magnetar flares (e.g.~\citealt{Wadiasingh2019}; \citealt{Cheng+20}); (4) a sufficiently high volumetric rate of magnetar birth to plausibly explain the observed FRB rate (e.g.~\citealt{Nicholl+17}), unlike other models involving rare cataclysmic events \citep{Ravi19}.

Despite these hints, several properties of the growing sample of FRBs appear$-$at least at first glance$-$to be in tension with magnetars as a primary source.  The first repeating source, FRB 121102 \citep{Spitler+16}, has been bursting nearly continuously (albeit interrupted by extended ``dark'' periods) for over 7 years; no known magnetar in our Galaxy matches this continuous level of activity.  One is forced to the conclusion that at least the most active repeating FRB sources arise from magnetars which are somehow different from the Galactic population, e.g. being of very young age \citep{Metzger+17,Beloborodov17}, formed via alternative channels than ordinary core-collapse supernovae  (CCSNe; \citealt{Metzger+17,Margalit+19,Zhong&Dai20}), or possessing other atypical property such as an unusually long rotational period \citep{Wadiasingh2019}.  

The recurrent fast radio burst FRB 180916 was recently shown by the CHIME/FRB collaboration to exhibit a 16 day period of unknown origin \citep{CHIME+20}.  Again, although known Galactic magnetars offer no clear explanation for periodic behavior at this scale (with the possible exception of candidate magnetar 1E 161348–5055 which has a measured period of 6.7 hr; \citealt{DeLuca2006}), reasonable variations in the properties of extragalactic magnetars (e.g. extremely slow rotation, precession, or presence in a binary) offer a potential explanation \citep{Lyutikov+20,Levin+20,Zanazzi&Lai20,Beniamini+20,Yang&Zou20,Tong+20}.  Furthermore, several FRBs have now been localized to host galaxies with low levels of star formation (\citealt{Bannister+19,Ravi+19}), uncharacteristic of the environments of magnetars in the Milky Way as being the product of relatively typical CCSNe.   

Perhaps most challenging to the magnetar model, until recently no Galactic magnetar had been observed to produce a radio burst with an energy matching those of known cosmological FRBs.  An FRB-scale burst was ruled from the giant magnetar flare in SGR 1806-20 insofar as it would have been detectable even in the side lobes of Parkes \citep{Tendulkar+16}.  Such behavior may still be consistent with some magnetar models, e.g. because of beaming of the radio emission or because the nature of the external environment (i.e. the {\it history} of the flaring activity) plays as important of a role in generating FRB emission as the flare itself \citep{Beloborodov17,Metzger+19,Beloborodov19}.  Nevertheless, skepticism regarding magnetar FRB models would only continue to grow if FRB-like emission were never seen in association with nearby, verifiable magnetars.

The observational situation changed abruptly with the recent discovery of a luminous millisecond radio burst from the Galactic magnetar SGR 1935+2154 \citep{Scholz_ATel13681,Bochenek_ATel13684}. The double-peaked burst, detected independently by CHIME \citep{Bandura+14} and STARE2 \citep{Bochenek+20}, was temporally coincident with an X-ray burst of significantly larger fluence \citep{Mereghetti_GCN27668,Zhang_ATel13687,Mereghetti+20}. This ``fast radio burst" is still a factor of $\sim 10$ less energetic than the weakest FRB previously detected from any localized cosmological FRB source.  It nevertheless represents an enormous stride in bridging the energy gap between Galactic magnetars and their hypothesized extragalactic brethren, providing new support to magnetar FRB models.

Here, we explore several implications of this discovery for the broader magnetar-FRB connection.  We emphasize that there exists no single ``magnetar model'', but rather a range of models which make drastically different predictions for the mechanism and location of the radio emission and the accompanying higher frequency radiation, some which this discovery lend credence to and others for which the model is placed in tension.

This paper is organized as follows.  In \S\ref{sec:observations} we summarize the observational picture regarding SGR 1935+2154 and its recent radio/X-ray activity.  In \S\ref{sec:model_independent_implications} we address several broad implications of this discovery in the context of magnetar models for cosmological FRBs.  In \S\ref{sec:magnetar_models} we discuss the implications of the coincident radio and X-ray flare from \source~for extant variations of the magnetar model.  Finally, we summarize and provide bulleted conclusions in $\S\ref{sec:conclusions}$.

\section{Summary of Observations}
\label{sec:observations}

\subsection{SGR 1935+2154}
\source~is a Galactic Soft Gamma Repeater (SGR) first discovered by \textit{Neil Gehrels Swift Observatory}'s Burst Alert Telescope (BAT) as a GRB candidate through a series of soft bursts from the Galactic plane \citep{Stamatikos+14, Lien+2014}. 
The source is associated with supernova remnant (SNR) G57.2+0.8 \citep{Gaensler14}. Distance estimates are uncertain, ranging from 6.6-12.5 kpc \citep{Sun11,Pavlovic+13,Kothes+18,Zhou+20,Mereghetti+20}, and throughout this paper we adopt a distance of $d = d_{10}\, 10 \, {\rm kpc}$.
Subsequent discovery of coherent X-ray pulsations of \source~with the \textit{Chandra X-ray Observatory} established the spin period of the magnetar, $P\simeq 3.2$ s \citep{Israel+14, Israel+16}. \textit{XMM-Newton} and \textit{NuSTAR} observations of the source during outburst in 2015 provided the magnetar's spin-down rate, $\dot{P}\simeq1.43\times10^{-11} \, {\rm s \, s}^{-1}$, which implies a surface dipolar magnetic field $B\simeq2.2\times10^{14} \, {\rm G}$, a spin-down luminosity  $L_{\rm sd}\simeq 1.7\times10^{34}$ erg s$^{-1}$, and characteristic spin-down age of $P/2\dot{P}\simeq3600$ years. 

We note, however, that the age estimate based on the SNR association, $\gtrsim 16$~kyr, is significantly older (\citealt{Zhou+20}; see also \citealt{Kothes+18}). This discrepancy between the dipolar and the SNR age estimates is similar to the one observed in the other magnetar associated with a SNR, Swift J1834.9--0846 (\citealt{Granot2017}; with a spin-down age of $4.9$~kyr and a SNR age between 5 and 100 kyr). Since the dipolar age estimate is expected to be inaccurate in case either the surface magnetic field evolves with time (e.g. \citealt{Colpi2000,Dall'Osso2012D,Beniamini2019}) or else the spin evolution is not dominated by dipolar radiation \citep{BT1998,Harding1999,Beniamini+20}, \cite{Granot2017} concluded that for Swift J1834.9--0846 the SNR age is likely to be the more realistic case. In this case, both Swift J1834.9--0846 and SGR 1935+2154 are significantly (by a factor of $\sim 4-10$) older than the majority of the observed Galactic magnetar population \citep{Beniamini2019}.

Radio observations with the Parkes telescope at $1.5 \, {\rm GHz}$ and $3 \, {\rm GHz}$ did not reveal any significant radio pulsations from \source~to a limiting flux of $0.1 \, {\rm mJy}$ and $0.07 \, {\rm mJy}$, respectively \citep{Burgay+14}. 
\textit{NCRA}/GMRT and \textit{ORT} observations at $\sim 300$ and $600$~MHz also found no radio pulses down to $0.4$, $0.2 \, {\rm mJy}$, respectively \citep{Surnis+14, Surnis+16}, followed by additional $14 \, \mu$Jy and $7 \mu$Jy limits by Arecibo (at $4.6$ and $1.4 \, {\rm GHz}$; \citealt{Younes+17}).

The non-detection of periodic radio pulses from the source has not impaired \source~from being a prolific X-ray burster. This magnetar has been active since its discovery with at least four outbursts on: July 5$^{\rm th}$ 2014; February 22$^{\rm nd}$ 2015; May 14$^{\rm th}$ 2016; and June 18$^{\rm th}$ 2016---each with an increasing number of bursts extending to higher energies \citep{Lin+20}.  During its active outburst cycles, \source~exhibited a remarkably bright burst on April 12$^{\rm th}$ 2015, detected by the four Interplanetary network (IPN) spacecrafts \citep{Kozlova+16}. This burst's long duration $\sim 1.7 \, {\rm s}$ and large fluence $\sim 2.5 \times 10^{-5}$ erg cm$^{-2}$ (radiated energy $\sim 10^{41}$ erg) place it in the rare class of ``intermediate'' SGR flares \citep{Mazets+99, Feroci+03, Mereghetti+09, gogus+11}.  

The trend that the bursts from SGR 1935+2154 have become progressively more energetic in the years since its initial discovery \citep{Younes+17} persists with the recent outburst, which is the most energetic yet \citep{Mereghetti+20,Zhang_ATel13687}. Similar behaviour has previously been observed for SGR~1806-20, which culminated with it's production of the most energetic magnetar giant flare seen to date (\citealt{Younes+15}).

\subsection{An ``FRB'' from SGR 1935+2154}

On April 10, 2020, a short soft X-ray burst was triangulated by the IPN to \source~\citep{Svinkin_GCN27527}. This was followed by a slew of bursts, extending to hard X-rays, detected over the following couple of weeks \citep{Veres_GCN27531,Ridnaia_GCN27554,Cherry_GCN27623,Hurley_GCN27625,Ridnaia_GCN27631,Palmer_ATel13675,Ricciarini_GCN27663,Marathe_GCN27664,Ridnaia_GCN27667,Mereghetti_GCN27668,Ridnaia_GCN27669,Lipunov_GCN27670,Younes_ATel13678,Kennea_ATel13679}.

On April 28, as part of this period of enhanced source activity, a bright millisecond radio burst, the first of its kind, was detected from \source~ \citep{Scholz_ATel13681}. 
The radio burst was associated with a short hard X-ray burst \citep{Mereghetti_GCN27668} peaking at energies $E_{\rm peak} \sim 70 \, {\rm keV}$ (\citealt{Zhang_ATel13687}; see also \citealt{Mereghetti+20}).
The detection by the CHIME/FRB backend in the 400--800 MHz band comprise two sub-burst components.
The bursts, each $\sim5$ ms wide and separated by $\sim30$ ms had a reported DM of $332.81 \, {\rm pc \, cm}^{-3}$.
This DM value is consistent with the observed $\approx 8.6$~s delay of the radio burst (at $400$ MHz) with respect to the peak of the X-ray counterpart flare as being almost entirely due to the cold plasma time delay \citep{Mereghetti_GCN27668,Zhang_ATel13696,Mereghetti+20}. 
An independent detection of the burst was also reported from the STARE2 radio feeds at the 1.4 GHz band \citep{Bochenek_ATel13684}. They report the burst arrival time and the DM value to be consistent with the CHIME detection, and constrain the peak fluence to be $>1.5 \, {\rm MJy \, ms}$. The much lower flux detected by CHIME versus STARE2 may be at least partly attributable to the fact that the burst was detected in the sidelobes of CHIME.

The compelling nature of this burst led to a search for track-like muon neutrino events with the IceCube observatory, with no significant neutrino signals detected along its direction \citep{Vandenbroucke_ATel13689}. Likewise, VLA followup of the source found no persistent or afterglow radio emission down to a flux of $\sim 50 \mu$Jy \citep{Ravi_ATel13690,Ravi_ATel13693}.

\begin{figure}
    \centering
    \includegraphics[width=0.45\textwidth]{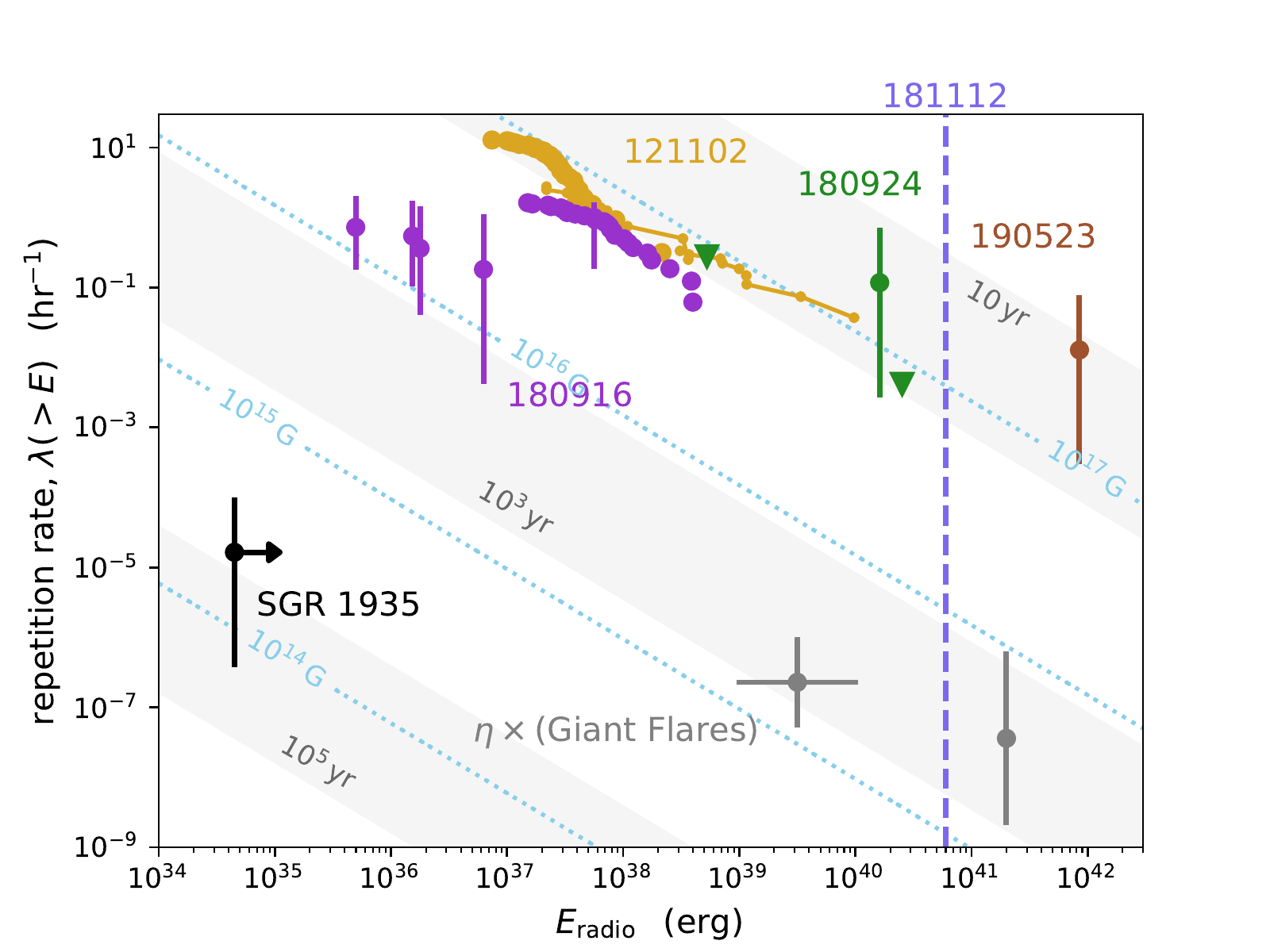}
    \caption{Repetition rate above a given emitted radio energy as a function of energy for \source~and localized FRBs: the repeating sources FRB 121102 \citep{Law+17,Gourdji+19} and  180916 \citep{Marcote+20,CHIME+20}, and apparently non-repeating FRBs 180924 \citep{Bannister+19}, 190523 \citep{Ravi+19}, and 181112 \citep{Prochaska+19}. 
    The energy of the recent radio burst from \source~is a factor of $\sim 10 d_{10}^2$ lower than the least energetic extragalactic FRB of known distance.  Applying the same ratio of radio to X-ray fluence measured for \source~ $\eta \sim 10^{-5}$ (eq.~\ref{eq:energy_ratio}) to giant magnetar flares would imply that Galactic magnetars are capable of powering even the most energetic cosmological FRBs.  However, a stark discrepancy exists between the activity (burst repetition rate) of Galactic magnetars and the sources of the recurring extragalactic FRBs (\S\ref{sec:model_independent_implications}). Scaling from the magnetic field and age of \source~implies that magnetar progenitors of extragalactic FRBs must have larger $B$-fields and younger ages (see \S\ref{sec:Rates} for further details).}
    \label{fig:luminosity_function}
\end{figure}

The millisecond-duration high-brightness temperature radio burst of \source~is unlike any other pulsar/magnetar phenomenology observed to date, with a luminosity exceeding that of even the most luminous Crab giant pulses (e.g.~\citealt{Mickaliger+12}) by several orders of magnitude.   Instead, the burst properties are suggestively similar to cosmological FRBs.  As pointed out by \cite{Bochenek_ATel13684}, placed at the distance of the nearest localized FRB~180916, $\simeq 149 \, {\rm Mpc}$ \citep{Marcote+20}, the \source~outburst would have been potentially detectable as a $> 7 \, {\rm mJy \, ms}$ burst, coming close to, albeit still lower than typical FRB fluences.
Stated energetically, SGR~1935's emitted radio energy is $E_{\rm radio} > 4 \times 10^{34} d_{10}^2 \, {\rm erg}$, within a factor of $10$ of the lowest-energy burst observed from any cosmological FRB of known distance to date, $\approx 5 \times 10^{35} \, {\rm erg}$ \citep{Marcote+20}. This is illustrated in Fig.~\ref{fig:luminosity_function}.  Also, at least from the standpoint of its X-ray fluence and duration, the ``FRB"-generating flare from \source~appears to be fairly typical among Galactic magnetar flares (Fig.~\ref{fig:burst_duration}; however, see discussion in $\S\ref{sec:model_independent_implications}$).  The immediate implication of all this is that magnetar activity akin to the burst observed from \source~should be contributing to the extragalactic FRB population.  

\section{SGR~1935 in the context of cosmological FRBs}
\label{sec:model_independent_implications}

In this section, we assume that all FRBs are produced by magnetar flares, with universal properties motivated by the SGR~1935 burst.  Proceeding under this strong assumption, we explore the implications for FRB energetics and repetition rates.  We are led to conclude that ``ordinary'' magnetars with activity-levels similar to SGR~1935 cannot alone explain the observed FRB population.

As discussed in \S\ref{sec:observations}, a contemporaneous X-ray flare was detected in coincidence with the radio burst of SGR~1935. The timing coincidence and similar sub-structure in both radio and X-ray bands implicates that the two be interpreted as counterparts \citep{Mereghetti_GCN27668,Zhang_ATel13696,Mereghetti+20}. The ratio between the radiated energy of the radio burst and its X-ray counterpart is
\begin{equation}
\label{eq:energy_ratio}
    \eta \equiv \frac{E_{\rm radio}}{E_{\rm X}} \sim 10^{-5} .
\end{equation}
Here we have calculated the X-ray burst energy $E_{\rm X} \approx 8 \times 10^{39} d_{10} \, {\rm erg}$ using the fluence $6.8 \times 10^{-7} \, {\rm erg \, cm}^{-2}$ reported by the  {\it Hard X-ray Modulation Telescope} (HXMT; \citealt{Zhang_ATel13687}). A similar fluence was reported by INTEGRAL \citep{Mereghetti+20}.  Likewise, we have estimated the radio energy $E_{\rm radio} \approx 4 \times 10^{34} d_{10} \, {\rm erg}$ by adopting the $1.5 \, {\rm MJy} \, {\rm ms}$ lower limit on the fluence reported by STARE2 \citep{Bochenek_ATel13684} and assuming a frequency width corresponding to the instrument bandwidth $BW \approx 250 \, {\rm MHz}$ \citep{Bochenek+20}.  The true value of $E_{\rm radio}$ (and hence $\eta$) may be somewhat larger than this estimate because the STARE2 radio fluence is quoted as a lower limit.  Furthermore, the fact that the same burst was detected also by CHIME at lower radio frequencies (albeit at a lower reported fluence, possibly attributable to the detection occurring in an instrumental sidelobe) suggests its spectral energy distribution is broadband, such that the true burst energy could be larger by a factor of $\gtrsim \nu/BW \approx 6$.

The low value of $\eta$ illustrates that magnetars are inefficient FRB producers.  Implications of this fact for specific magnetar FRB models are discussed later (\S\ref{sec:magnetar_models}).  Regardless of the emission mechanism, the active lifetime of cosmological recurrent FRB sources cannot be long if FRB  emission is similarly inefficient for such sources.  For the activity rate and radio fluences of FRB~121102 (e.g.~\citealt{Law+17}), the radio-inefficiency $\eta \sim 10^{-5}$ implies that the FRB-generating engine must be losing energy at a rate of $\sim$several $\times 10^{39} \, {\rm erg \, s}^{-1}$ (itself only a lower-limit if the luminosity-function is energetically-dominated by low energy undetectable bursts; \citealt{Gourdji+19}).  For FRB~180916 the repetition rate and luminosity function point to qualitatively similar requirements on the power output of the central engine, $\gtrsim 5 \times 10^{38} \, {\rm erg \, s}^{-1}$ \citep{CHIME+20}. 

If recurrent FRBs are powered by magnetars, then their active lifetime is at the very least limited by their total magnetic energy reservoir $E_{\rm mag} \sim 3 \times 10^{49} \, {\rm erg} \, (B / 10^{16})^2$,
\begin{equation}
    \tau_{\rm active} \sim \frac{E_{\rm mag}}{\dot{E}_{\rm FRB} / \eta} \sim 200 \, {\rm yr} 
    \, \left(\frac{B}{10^{16} \, {\rm G}}\right)^2 \left(\frac{\eta}{10^{-5}}\right) ,
    \label{eq:tactive}
\end{equation}
where $B$ is the interior magnetic field strength and we have taken $\dot{E}_{\rm FRB} \sim 5 \times 10^{34} \, {\rm erg \, s}^{-1}$ motivated by FRB~121102 \citep{Law+17}. 
Even for large interior fields $B \gtrsim 10^{16} \, {\rm G}$, the maximum active lifetime is significantly shorter than the $16 \, {\rm kyr}$ estimated age of \source~\citep{Zhou+20}, or indeed of any other known Galactic magnetar.

Based purely on their X-ray behavior, magnetars as active as FRB~121102 or other repeating FRB sources do not exist in our own Galaxy. These points suggest that if cosmological FRBs originate from magnetar progenitors, at least a subset of these magnetars must be far more active than \source, and are perhaps formed via different mechanisms than Galactic magnetars \citep{Margalit+19}.  We further quantify this point in the next section by calculating the extragalactic detection rate of \source-like events. 

\begin{figure}
    \centering
    \includegraphics[width=0.5\textwidth]{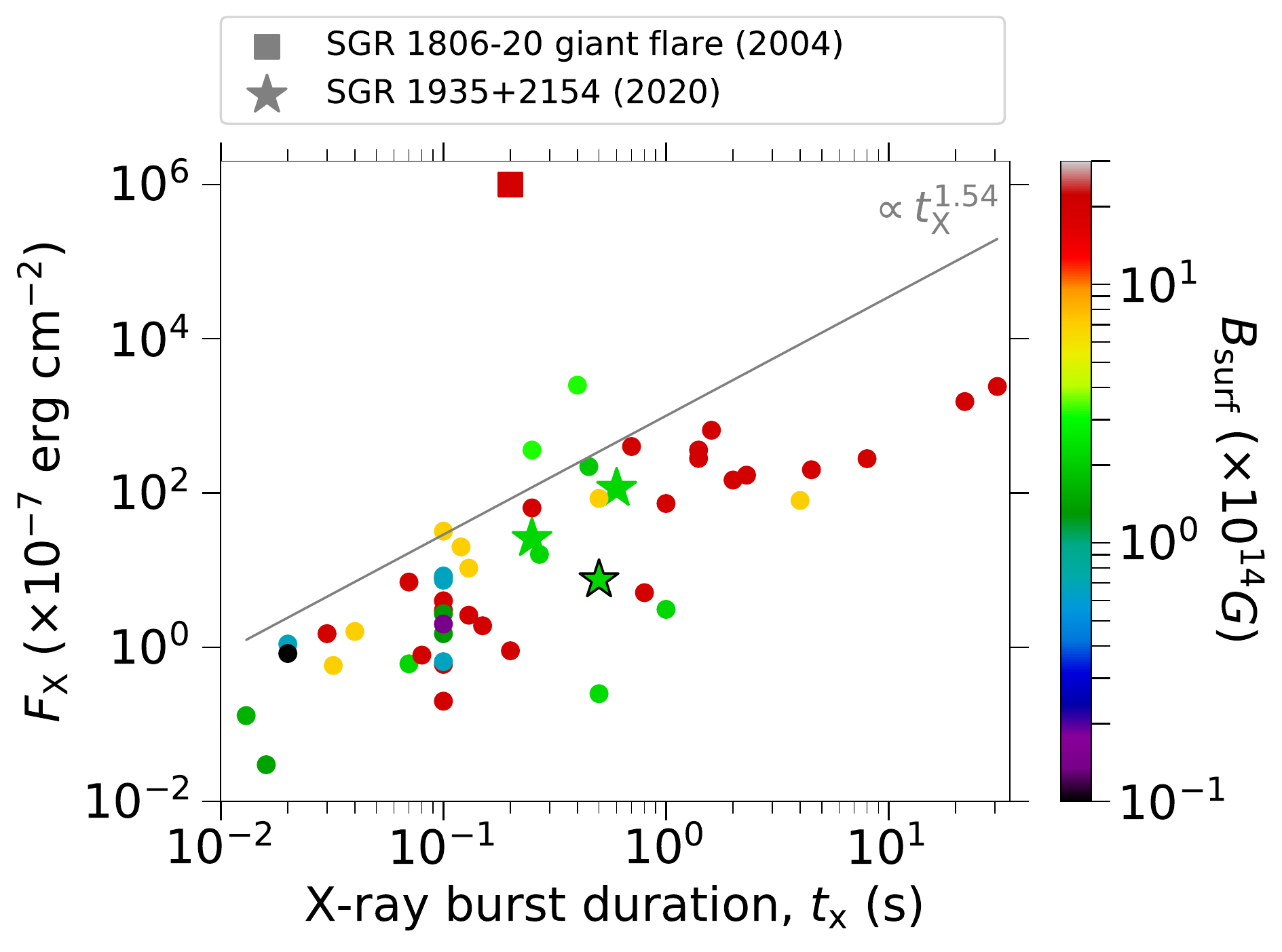}
    \caption{
    X-ray fluence and duration of Galactic magnetar flares. The \source~flare associated with the April 28th radio burst (green star, black border) is typical amongst Galactic magnetar flares.
    The grey line shows the fluence--duration correlation found by \citet{Gavriil+04}, $F_{X} \propto t_{X}^{1.54}$.
    Points are colored by the surface magnetic field of each magnetar (the McGill magnetar catalog$^{\dagger}$; \citealt{OlausenKaspi14}).  In the event of a prolonged outburst ($1\sim3$ months), we show only the brightest burst of a given outburst. Likewise, for bursts clustered over timescales of a few seconds, we show the fluence corresponding to that of the initial peak, and not of the entire envelope of bursts. 
    The anomalous concentration of bursts with $t_{\rm X} \sim 0.1 \, {\rm s}$ can be attributed to 
    instrument temporal resolution. 
    \\ \noindent $^{\dagger}$\url{http://www.physics.mcgill.ca/~pulsar/magnetar/main.html}}
    \label{fig:burst_duration}
\end{figure}
\begin{figure}
    \centering
    \includegraphics[width=0.5\textwidth]{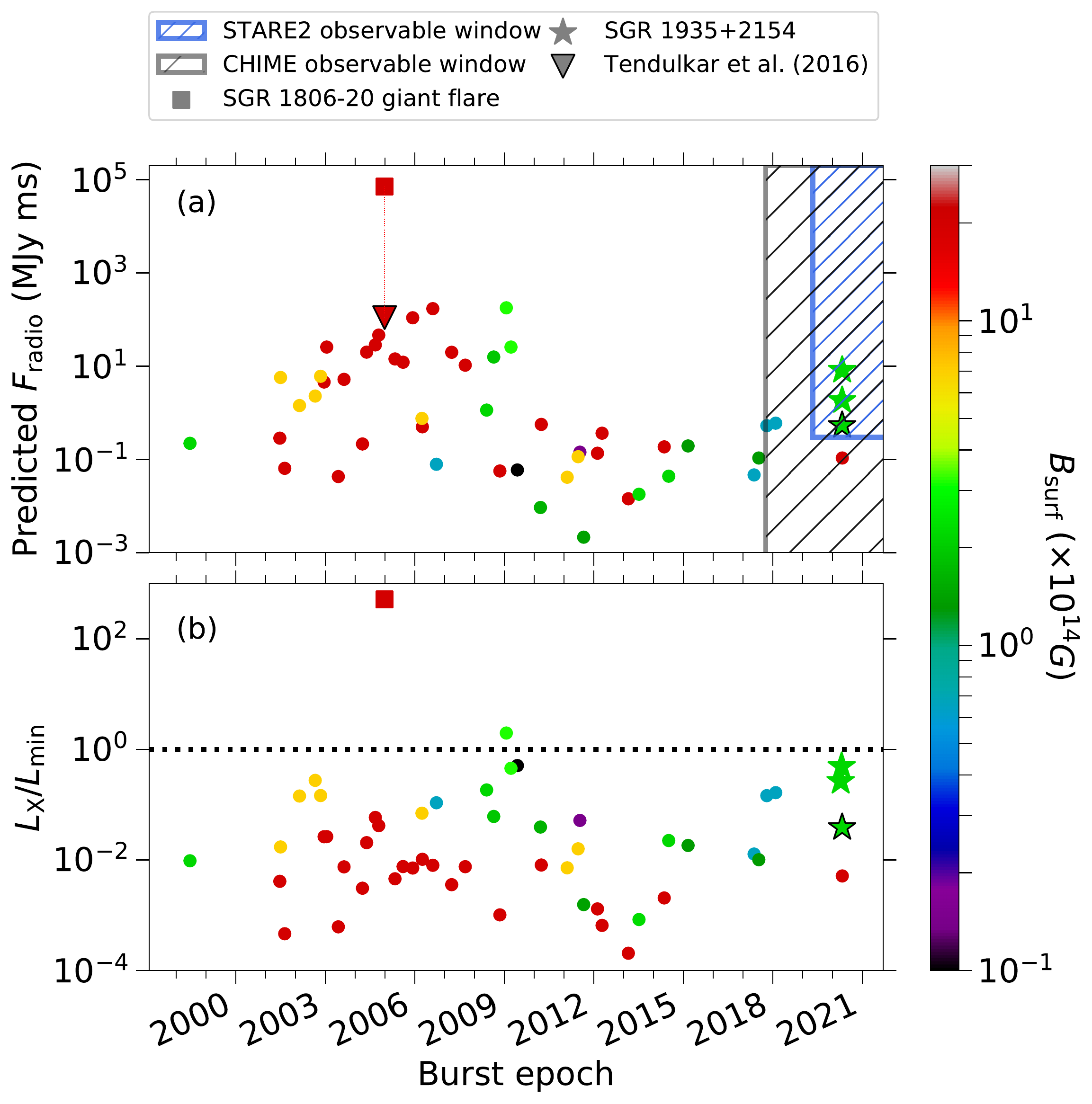}
    \caption{Properties of Galactic magnetar flares as a function of burst epoch. Different markers denote the same set of sources described in Fig.~\ref{fig:burst_duration}. (a) {\it Top Panel:} the predicted fluence of radio burst counterparts to magnetar flares, assuming a radio-to-X-ray efficiency $\eta\sim 10^{-5}$ (eq.~\ref{eq:energy_ratio}). Hatched blue and grey regions denote the observable windows of STARE2 and CHIME/FRB, respectively (assuming on-beam sensitivity limits). The upper limit on radio emission 
    from the SGR~1806--20 giant flare (red square) is shown as an upside-down triangle \citep{Tendulkar+16}. (b) {\it Bottom Panel:} the luminosity of the same bursts relative to the magnetic Eddington luminosity $L_{\rm min}$ (\citealt{Paczynski1992}; see also \S\ref{sec:model_independent_implications}), a rough scale for the minimum luminosity burst that is capable of driving baryonic outflows via radiation pressure.}
    \label{fig:burst_epoch}
\end{figure}

Before turning to extragalactic sources, we examine more closely the burst from \source~in light of other bursts from Galactic magnetars observed in the past $\sim 20$ years. Figure \ref{fig:burst_duration} depicts the X-ray fluence ($F_{\rm X}$) and the duration ($t_{\rm X}$) of the recent X-ray bursts from \source, alongside other bright bursts\footnote{The duration and fluence are obtained from an extensive search of NASA GCN circulars archive (\url{https://gcn.gsfc.nasa.gov/gcn3_archive.html}), and The Astronomers' Telegram (\url{http://astronomerstelegram.org/}).}.  Here, the duration of the burst ($t_{\rm X}$) is defined as the interval of time when $5\%$ and $95\%$ of the total background subtracted counts are recorded \citep{Gogus01}. A rough correlation is seen between the fluence and duration \citep{Gogus01}, approximately satisfying $F_X\propto t_{\rm X}^{1.54}$ consistent with reports in the magnetar literature for the bursts from single sources \citep{Gavriil+04}. Although the bursts from \source~are bright, they fit within this trend, and are not exceptional in terms of their fluence or overall-envelope duration with respect to bright bursts from other Galactic magnetars.
However, the burst associated with \source's radio emission may have exhibited a harder spectrum than other bursts \citep{Mereghetti+20}, indicating that perhaps this particular burst was produced via a different emission mechanism, and that this mechanism may be related to FRB production. In $\S\ref{sec:MMS_model}$ we present a scenario in which shock-powered X-rays are generated concurrently with the coherent radio emission; however, as thermal X-rays may be produced in the magnetosphere during flaring by other mechanisms, it is plausible that multiple sources of X-ray emission contribute to different flares, or within a single burst. Nonetheless, this raises the question of whether other FRBs from Galactic magnetars should have already been observed in the past?

Assuming the value of $\eta$ from the recent radio detection of \source~is universal across all flares, we predict the fluence of the radio emission ($F_{\rm radio}$) that should accompany the X-ray bursts from a large population of Galactic magnetar outbursts over the past $\sim2$ decades (the same sample shown in Fig.~\ref{fig:burst_duration}). Figure \ref{fig:burst_epoch}(a) shows $F_{\rm radio}$ as a function of the burst date, with reference to the estimated sensitivity (hatched regions) of STARE2 and CHIME/FRB telescopes. We see that two flares preceded the April 28$^{\rm th}$ FRB event from \source~with higher predicted $F_{\rm radio}$---one on April 10$^{\rm th}$ 2020 \citep{Veres_GCN27531} and another on April 22$^{\rm nd}$ 2020 \citep{Ridnaia_GCN27631}. These events fall within the nominal STARE2 observable window, despite no reported radio detection. We note that relatively few magnetar flares have occurred over the last few years, when STARE2 and CHIME have been operational, at a fluence that would have been detectable by these instruments.  Also note that the on-beam CHIME/FRB sensitivity shown is not the relevant one for most bursts, which will occur in the sidelobes of the telescope.

In the exceptional case of the giant flare from SGR~1806--20 \citep{Palmer+05,Hurley+05}, the upper limit on radio emission (\citealt{Tendulkar+16}; estimated based on sidelobe sensitivity of Parkes) is smaller than the predicted radio fluence (for $\eta\sim 10^{-5}$) by a factor of $\sim 100$ (Fig.~\ref{fig:burst_epoch}; red triangle and square, respectively). As we discuss later in \S\ref{sec:MMS_model}, this may be understood in some magnetar FRB models, for different beaming properties of the radio emission relative to the higher frequency counterpart in giant flares as opposed to less energetic magnetar flares. More generally, the fact that one FRB was observed out of only a few magnetar bursts where we might have expected a detection, suggests that the FRB beaming is rather modest. The beaming can be directly probed in the future, once more X-ray bursts from \source~ with comparable luminosities are observed, by searching for a possible correlation between FRB detectability and the rotational phase of the magnetar. 
  
Using the X-ray burst fluence for each source, and distance estimates from the McGill magnetar catalogue \citep{OlausenKaspi14}\footnote{\url{http://www.physics.mcgill.ca/~pulsar/magnetar/main.html}}, we estimate the intrinsic burst X-ray luminosity ($L_{\rm X}$).  For the selected bursts, we depict in Figure \ref{fig:burst_epoch} the value of $L_{\rm X}$ normalized to the individual source's ``magnetic Eddington luminosity" $L_{\rm min}=3.5\times 10^{38} (B_{\rm surf} / 10^{12}\,{\rm G})^{4/3}\mbox{erg s}^{-1}$ \citep{Paczynski1992}, where $B_{\rm surf}$ is the surface dipole magnetic field strength. $L_{\rm min}$ is a rough scale for the minimum luminosity of a burst that can drive a mass outflow via radiation pressure.  

Note that the FRB-producing flare from \source~obeyed $L_{\rm X} < L_{\rm min}$ (black-outlined green star in Fig.~\ref{fig:burst_epoch}) and hence would not have necessarily been expected to produce a baryon-loaded outflow, at least via radiation pressure.  This is potentially relevant because some FRB models$-$particularly the baryonic shock models (\S\ref{sec:MMS_model})$-$require an upstream medium into which the shocks collide.  On the other hand, other mechanisms than radiation pressure (e.g. magnetic stresses) may also play a role in mass ejection, and the quantity of mass in the external medium needed in the synchrotron maser scenario is extremely modest.  

\subsection{Rates: A Single Magnetar Population?}
\label{sec:Rates}

Given the $3.6$ steradian field-of-view of STARE2 \citep{Bochenek+20} and the fact that a single magnetar radio burst has been detected during the $\sim 300$~day operating period of the experiment, we estimate the rate of \source-like magnetar radio bursts (taking the number of active magnetars in the Galaxy to be $N=29$; \citealt{OlausenKaspi14}) to be 
$\lambda_{\rm mag} \in [0.36,80] \times 10^{-2}$ per magnetar per year at 95\% confidence (assuming Poisson statistics; \citealt{Gehrels86}).

The radio-burst activity (repetition) rate of SGR~1935 estimated above is plotted in Fig.~\ref{fig:luminosity_function} in comparison to cosmological FRBs. We show here the full sample of published localized FRB sources, where the radio-emitted (isotropic-equivalent) energy can be reliably calculated:
repeating FRB~121102 (yellow; \citealt{Law+17,Gourdji+19}), the recently localized CHIME repeater FRB~180916 (purple; \citealt{Marcote+20,CHIME+20}), and the apparently non-repeating FRBs~180924 (green, upper limits denoted by upside-down triangles; \citealt{Bannister+19}), 190523 (brown; \citealt{Ravi+19}), and 181112 (blue; \citealt{Prochaska+19}). We have used quoted rates where possible \citep{Law+17,Gourdji+19,CHIME+20} and otherwise estimated Poissonian rates based on quoted field exposure times, where available.
Grey markers with errorbars show the repetition rate (from \citealt{Tendulkar+16}) and radio energy implied if some giant magnetar flares produce radio emission with the same efficiency $\eta \sim 10^{-5}$ between X-ray and radio fluences. This indicates that Galactic magnetars may be energetically capable of powering even the most luminous FRBs, though for one particular flare from SGR~1806-20 such radio emission is ruled out \citep{Tendulkar+16}.
Finally, we note that radio pulses observed from M87 by \cite{Linscott1980} $\sim 40$yr ago, though unconfirmed by subsequent followup, may have been the earliest recorded FRB detections, and correspond to $E_{\rm radio} \sim 10^{38} \, {\rm erg}$ on this figure.

Figure~\ref{fig:luminosity_function} shows that, at comparable energy, the SGR~1935 radio burst is $\sim 10^{-5}$ less frequent than FRB~180916 bursts. Within the hypothesis that magnetars are also the progenitors of such cosmological FRB, this stark discrepancy in activity rate may be attributable to the magnetar age and internal field strength. Indeed, \cite{Margalit+19} show that magnetar activity scales strongly with magnetic field, $\dot{E} \propto B^{3.2}$ \citep{Dall'Osso2012D,BeloborodovLi16}.\footnote{Also note that activity may increase with the NS {\it mass}, because at sufficiently high central densities the star can cool through direct URCA reactions, which hastens the release of magnetic energy from the core \citep{BeloborodovLi16}.}  
This is shown schematically with the dotted-blue contours in Fig.~\ref{fig:luminosity_function}, assuming that $\lambda(>E)E \sim \dot{E} \propto B^{3.2}$, and scaling the $B$-field from $\sim 2 \times 10^{14} \, {\rm G}$, the external dipole field of \source~(note that the internal field is what sets $\dot{E}$, and thus we implicitly assume that the internal field is comparable to the external dipole field for this object).
This assumed $B$-field implies an active lifetime of $\sim 70 \, {\rm kyr}$ for SGR~1935 \citep[e.g.][their eq.~1]{Margalit+19}, consistent with the $\gtrsim 16 \, {\rm kyr}$ source age. 
Contours of active lifetime ($\sim$magnetar age) are also shown in Fig.~\ref{fig:luminosity_function} (grey shaded regions), scaling from SGR~1935's estimated age as $\tau_{\rm active} \propto B^{-1.2}$ \citep{Dall'Osso2012D}.
The $5$ order of magnitude higher repetition rate of FRB~180916 would thus imply $B_{\rm 180916} \sim (10^{5})^{1/3.2} B_{\rm SGR1935} \sim 10^{16} \, {\rm G}$, in agreement with separate lines of argument, e.g. requirements for FRB~180916 periodicity to be attributable to magnetar precession \citep{Levin+20}.

An alternative possibility is that the repetition rate increases significantly with the periodicity \citep{Wadiasingh2020}. In this case, the periodicity of 180916 (and its high activity relative to SGR 1935+2154) may be ascribed to an extremely long period magnetar \citep{Beniamini+20}. Interestingly, this option too, requires a large internal magnetic field, $> 10^{16} \, {\rm G}$, at birth (see \citealt{Beniamini+20} for details).

The maximum distance up to which \source's radio flare would be detectable by typical FRB search facilities sensitive to $F_{\rm lim} \sim 1 \, {\rm Jy \, ms}$ fluence radio pulses is $D_{\rm lim} \approx 12 \, {\rm Mpc} \, d_{10} (F_{\rm lim} / 1 \, {\rm Jy \, ms})^{-1/2}$.
The birth rate of ``ordinary'' Galactic magnetars in the local Universe can be estimated as a fraction $f_{\rm CCSN} \approx 0.1-1$ \citep{Beniamini2019} of of the CCSN rate, $\Gamma_{\rm CCSN} \approx (0.71 \pm 0.15) \times 10^{-4} \, {\rm Mpc}^{-3} \, {\rm yr}^{-1}$ \citep{Li+11}. Thus, the rate of potentially-detectable FRBs produced by \source-like bursts is
\begin{eqnarray}
\label{eq:rate_basic}
    \mathcal{R}(>F_{\rm lim}) 
    &\approx& \frac{4 \pi}{3} D_{\rm lim}^3 f_{\rm CCSN} \Gamma_{\rm CCSN} \tau_{\rm active} \lambda_{\rm mag}
    \\ \nonumber
    &\in& \left[ 0.004, 1.4 \right] \, {\rm sky}^{-1} \, {\rm day}^{-1} \, 
    \\ \nonumber
    &&\times d_{10}^3 \left(\frac{f_{\rm CCSN}}{0.1}\right) \left(\frac{\tau_{\rm active}}{10^4 \, {\rm yr}}\right) \left(\frac{F_{\rm lim}}{1 \, {\rm Jy \, ms}}\right)^{-3/2} .
\end{eqnarray}
This is much lower than the Parkes estimated FRB rate $\approx 1.7^{+1.5}_{-0.9} \times 10^3 \, {\rm sky}^{-1} \, {\rm day}^{-1}$ above a fluence of $2 \, {\rm Jy} \, {\rm ms}$ \citep{Bhandari+18}. 

The estimate above shows that magnetars bursting at the same rate {\it and energy} as the April 28$^{\rm th}$ \source~radio burst cannot contribute noticeably to the FRB population. 
However, clearly magnetar flares span a range of energies (Fig.~\ref{fig:burst_duration}), making it natural to ask whether \source-like magnetars can reproduce the FRB rate if one extrapolates radio production with a similar efficiency to more energetic magnetar flares.

We therefore calculate the all-sky rate of FRBs above a limiting fluence threshold assuming that all magnetars repeat following a luminosity-distribution $\lambda_{\rm mag}(>E) \propto (E/E_{\rm min})^{-\gamma}$ for burst energies between $E_{\rm min}$ and $E_{\rm max}$. We set $E_{\rm max} = \eta E_{\rm mag} \propto B^2$ dictated by the magnetic energy reservoir of the magnetar (this is consistent with the energies of the three observed Galactic giant flares relative to the magnetic dipole energies of the magnetars producing them, see e.g. \citealt{Tanaka2007}),
and fix the minimum energy to that observed for the April 28$^{\rm th}$ \source~flare\footnote{Given that the luminosity function we have adopted dictates fewer bursts at higher energies (for $\gamma>0$), it is statistically unlikely that the detected radio burst of \source~exceed $E_{\rm min}$ significantly.}.
The magnetar birth-rate $\Gamma_{\rm birth}(z)$ is assumed to follow the cosmic star-formation rate \citep{HopkinsBeacom06} and is anchored to 10\% the CCSN rate at $z=0$ \citep{Li+11}. The integrated rate is thus
\begin{eqnarray}
\label{eq:rate}
    \mathcal{R}(>F_{\rm lim}) 
    &=& \int dV(z) \Gamma_{\rm birth}(z) \tau_{\rm active} 
    \\ \nonumber
    &\times& \int dE \lambda_{\rm mag}(E) \Theta \left[ E - 4\pi D_{\rm L}(z)^2 F_{\rm lim} \right] .
\end{eqnarray}

\begin{figure}
    \centering
    \includegraphics[width=0.45\textwidth]{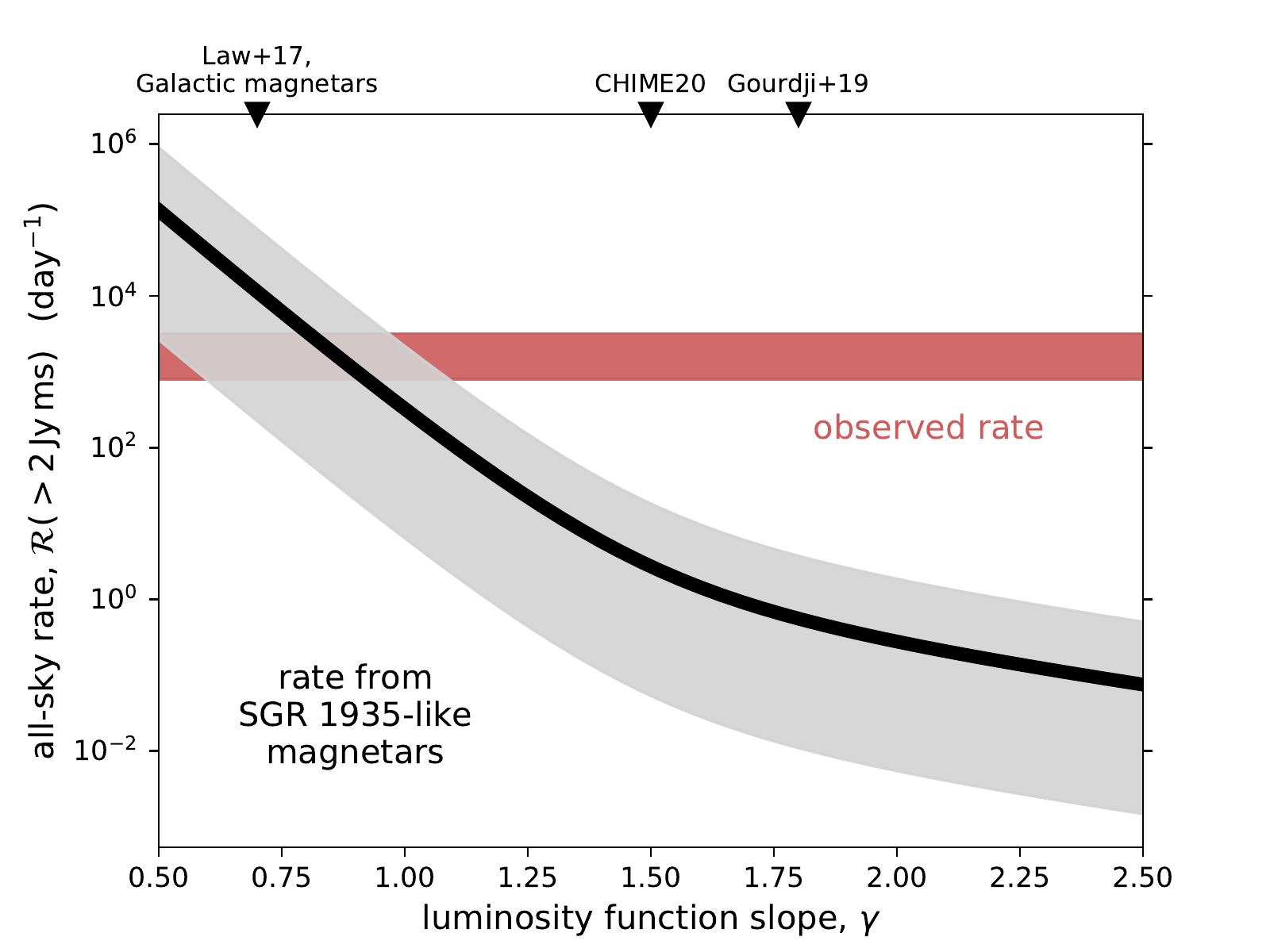}
    \caption{All-sky rate of FRBs above a limiting fluence of $2 \, {\rm Jy \, ms}$. The black curve (and shaded uncertainty region) shows the FRB rate that would be detected from a single population of ``normal'' magnetars (assumed to be born at 10\% the CCSN rate) with active lifetimes and radio repetition-rates set by \source. The repetition rate at $E_{\rm min} = 2 \times 10^{34} \, {\rm erg}$ is fixed to the inferred rate of \source~bursts at this energy, and is extrapolated to higher-energies by a luminosity-function $\lambda_{\rm mag}(>E) \propto E^{-\gamma}$. For large $\gamma$ the all-sky rate is dominated by weak, common bursts at $\sim E_{\rm min}$ and the rate falls short of the observed FRB rate (horizontal red; \citealt{Bhandari+18}) by several orders of magnitude (eq.~\ref{eq:rate_basic}). However, for $\gamma \lesssim 1$ the rate of FRBs from \source-like sources (extrapolated to higher energies) accommodates and even over-produces the observed FRB rate.  The values of $\gamma$ found by different studies are indicated on the top axis, including of the X-ray/gamma-ray luminosity function of Galactic magnetars.}
    \label{fig:FRB_rate}
\end{figure}

We set $\tau_{\rm active} = 24 \, {\rm kyr}$ and fix $\lambda(>E_{\rm min})$ to the estimated rate of SGR~1935 radio bursts. 
The former is determined from the minimum SNR age estimated by \cite{Zhou+20} and scaled to our adopted distance of $10 \, {\rm kpc}$.
We then integrate over cosmological redshift and calculate the rate as a function of the slope of the luminosity-function, $\gamma$.
Fig.~\ref{fig:FRB_rate} shows the resulting all-sky rate as a function of $\gamma$ (the only free variable in the calculation). For large $\gamma$, the rate is dominated by bursts of energy $\sim E_{\rm min}$ and we recover the result of eq.~(\ref{eq:rate_basic}), namely that the rate of Galactic magnetars is too low compared to cosmological FRBs. However, for $\gamma < 1.5$ the rate becomes dominated by the high-energy tail of the luminosity function, and the number of detectable magnetar radio bursts increases. In particular, we find that, for $\gamma \lesssim 1$---magnetars of a similar kind to \source~can account for the entire FRB rate.

Although extending the luminosity function of radio bursts to higher energies can allow (for certain values of $\gamma$) ``ordinary'' magnetars similar to \source~to reproduce the number of observed FRBs, this scenario falls short of explaining many other observables.
In particular, and as discussed previously---the per-source activity rate for such a population would be far too low to explain prolific repeaters like FRBs 121102 and 180926. For the same reason, most FRB sources would be detected only once and the relative number of repeaters versus apparently-non-repeaters would be very small. The scenario would also predict average FRB source distances that are far nearer than localized sources or DM-estimated distances.

\subsection{Two-Component Population}

To further explore the requirements on possible magnetar progenitors of cosmological FRBs, we extend the calculation of the previous section and model a two-component magnetar population as necessitated by the observations: 
magnetars with low activity levels like \source, and magnetars that are very active.
A natural question this will allow us to address is whether the active population is consistent with an earlier evolutionary stage (a younger version) of the same \source-like population, or whether one requires a distinct population altogether (e.g. a rare subset of magnetars born through unique channels).

We again calculate the all-sky FRB rate using eq.~(\ref{eq:rate}), accounting for the two populations contributing to FRB production. As before, the only free parameter describing the ``ordinary magnetar'' population is the luminosity-function slope $\gamma$. The second, ``active'', population is then scaled from the former population as a function of the internal magnetic field $B$ and relative birth-rate $f_{\rm CCSN}$.\footnote{We assume the ``ordinary-magnetar'' population is born at 10\% the CCSN rate, independent of the fractional birth rate $f_{\rm CCSN}$ of the ``active'' population.}

The magnetic field of the magnetar enters in setting $E_{\rm max} \propto B^2$, the active lifetime of sources $\tau_{\rm active} \propto B^{-1.2}$ \citep{Dall'Osso2012D,BeloborodovLi16}, and the repetition-rate $\lambda_{\rm mag}(>E_{\rm min})$. The latter is proportional to 
$\propto B^{3.2}$ for $\gamma>1$ and $\propto B^{1.2+2\gamma}$ for $\gamma<1$.\footnote{This is calculated based on the magnetic-field scaling of $E_{\rm max}$ and $E_{\rm min}$, the fact that $\dot{E} \propto \lambda_{\rm mag}(>E_{\rm min}) E_{\rm min}$ for $\gamma>1$ and $\lambda_{\rm mag}(>E_{\rm max}) E_{\rm max}$ for $\gamma>1$, and that $\dot{E} \propto B^{3.2}$; \citep{Dall'Osso2012D,BeloborodovLi16,Margalit+19}.} 
We normalized $\tau_{\rm active}$ and $\lambda_{\rm mag}(E_{\rm min})$ at the magnetic field of \source~($\simeq 2.2 \times 10^{14} \, {\rm G}$\footnote{This value is based on the inferred external dipole field of \source. The internal field may in fact be larger.}) to the age ($\gtrsim 24 d_{10} \, {\rm kyr}$) and radio repetition-rate of SGR~1935.
We then integrate eq.~(\ref{eq:rate}) and compare the resulting rate, a function of the three free parameters $\{B,\gamma,f_{\rm CCSN}\}$, to the observed FRB rate \citep{Bhandari+18}.
From the point of view of the all-sky rate alone, there exists a degeneracy between the number of FRB sources in the Universe ($\propto f_{\rm CCSN}$) and the repetition-rate of each source (a function of $B$).
However, this degeneracy can be broken by constraining the observed repetition rate of prolific repeaters.

\begin{figure}
    \centering
    \includegraphics[width=0.45\textwidth]{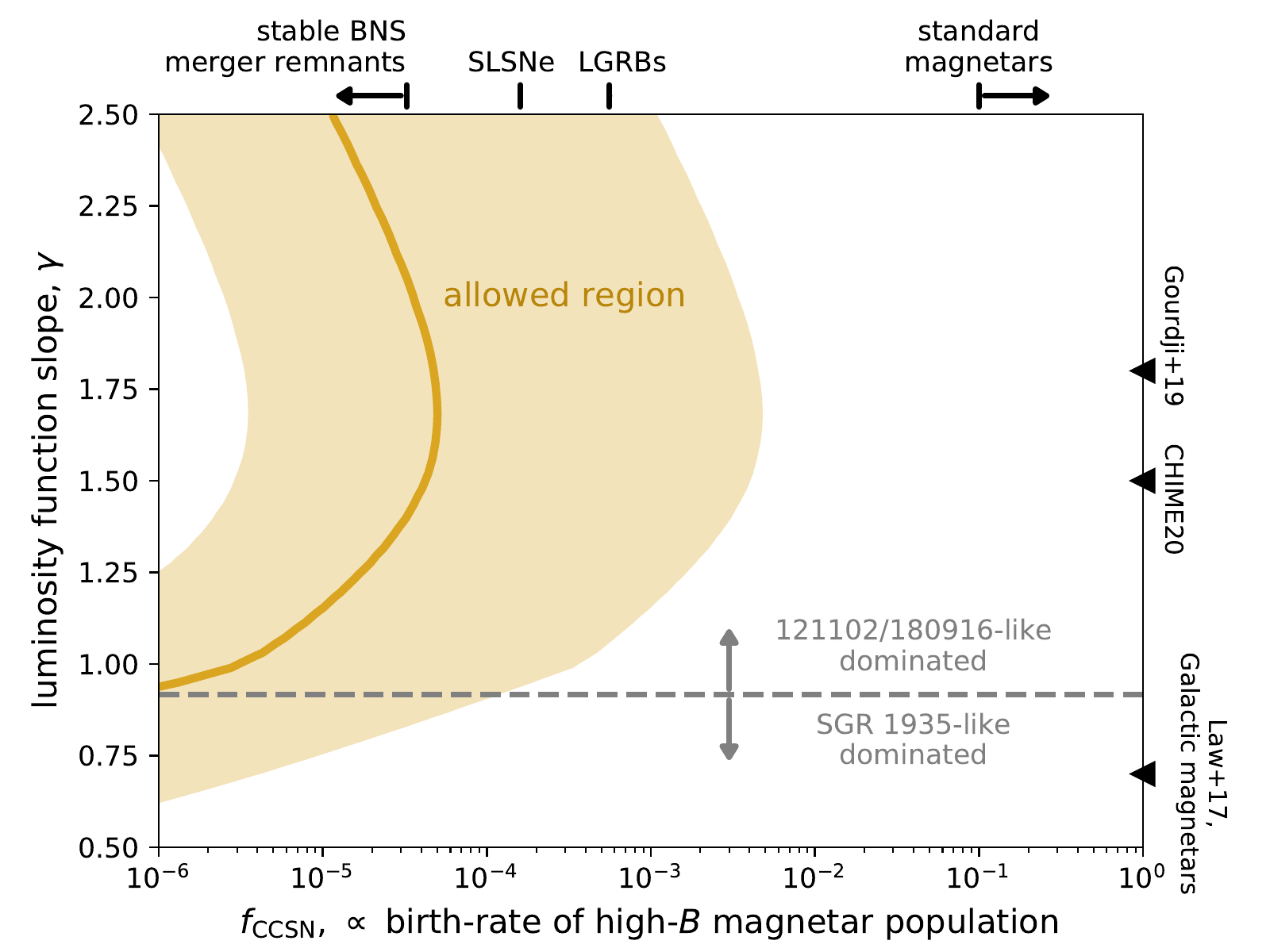}
    \caption{A two-component magnetar population consisting of: (i) ``ordinary'' magnetars fixed to properties of \source~and extrapolated as a function of the luminosity-function slope $\gamma$, and (ii) a magnetar population whose activity rate (parameterized via the internal magnetic field) and birth rate (a fraction $f_{\rm CCSN}$ of the CCSN rate) are allowed to vary (the value of $\gamma$ is identical to both populations). The yellow curve and shaded uncertainty region shows the allowed parameter space constrained by both the all-sky FRB rate and the per-source activity (repetition rate). The latter constraint forces $f_{\rm CCSN} \ll 1$, indicating that the second, ``active'' population of magnetars, must be volumetrically rare. This rules out the hypothesis that active cosmological repeating FRBs are younger versions of all \source-like magnetars (for which $f_{\rm CCSN} \gtrsim 0.1$).}
    \label{fig:repetition_rate}
\end{figure}

Figure~\ref{fig:repetition_rate} shows, for values of the magnetic field $B(\gamma,f_{\rm CCSN})$ required to fit the observed FRB rate (and its uncertainties), the region in $\{\gamma,f_{\rm CCSN}\}$ parameter-space where the average repetition rate of the ``active magnetar'' population equals the observed repetition-rate of FRBs 121102 and 180916. For this, we take a fiducial value $\lambda_{\rm mag}(>10^{38}\,{\rm erg}) = 1 \, {\rm hr}^{-1}$ (see Fig.~\ref{fig:luminosity_function}) however the uncertainties encompass a few orders-of-magnitude leeway in this assumed value.
On the basis of Fig.~\ref{fig:repetition_rate} one can conclude that, regardless of the luminosity-function slope $\gamma$, fitting both the all-sky FRB rate and the activity level of repeating FRBs requires a rare class of progenitors, $f_{\rm CCSN} \ll 1$.
This is in line with many previous studies \cite[e.g.][]{Nicholl+17}, and here we have extended these by adding the, possibly inevitable (though likely subdominant), contribution of a second population of ``normal magnetars'', and utilizing scaling-relations anchored to the new observations of \source.

The fact that $f_{\rm CCSN} \ll 1$ is required of the ``active population'' implies that these sources cannot be interpreted as younger-incarnations of the \source-like magnetar population as a whole, since this would require the same birth-rate for both populations, i.e. $f_{\rm CCSN} \sim 0.1$.
In the above we have assumed isotropic radio emission, as we expect beaming to be modest at most (\S\ref{sec:model_independent_implications}). If radio emission is beamed with some preferential directionality (as in pulsars) then the number of FRB emitting magnetars in the Universe may be larger than implied by the observed population. By ``hiding'' FRB sources this way, $f_{\rm CCSN}$ may be pushed to larger values. We note however that if bursts are instead beamed towards random directions, then the source birth rate as estimated above will be largely unchanged.

To further compare the resulting population against observational constraints, we calculate the expected number of repeating, and apparently-non-repeating, FRB sources detected by a mock survey of this population. We assume a limiting fluence of $F_{\rm lim} = 4 \, {\rm Jy \, \rm ms}$ and average repeat-field exposure time of $T_{\rm exp} = 40 \, {\rm hrs}$, parameters motivated by the CHIME FRB survey. We then calculate the (Poissonian) number of sources for which only a single burst would be detected,
\begin{equation}
\label{eq:Nnonrep}
    N_{\rm non-rep} 
    = \int dV(z) f_{\rm mag} \Gamma_{\rm birth}(z) \tau_{\rm active} \mu e^{-\mu}
\end{equation}
where 
\begin{equation}
    \mu(z) \equiv \lambda_{\rm mag}\left(>4\pi D_{\rm L}(z)^2 F_{\rm lim}\right) T_{\rm exp} ,
\end{equation}
and summing the contribution from both active and \source-like populations.
The number of sources classified as repeaters by the same survey is similarly calculated as
\begin{equation}
\label{eq:Nrep}
    N_{\rm rep}
    = \int dV(z) f_{\rm mag} \Gamma_{\rm birth}(z) \tau_{\rm active} 
    \\ \nonumber 
    \left[ 1 - \frac{1}{2} \Gamma \left( 2, \mu \right) \right] 
\end{equation}
where $\Gamma(2,\mu)$ is the incomplete gamma-function.

\begin{figure*}
    \centering
    \includegraphics[width=0.45\textwidth]{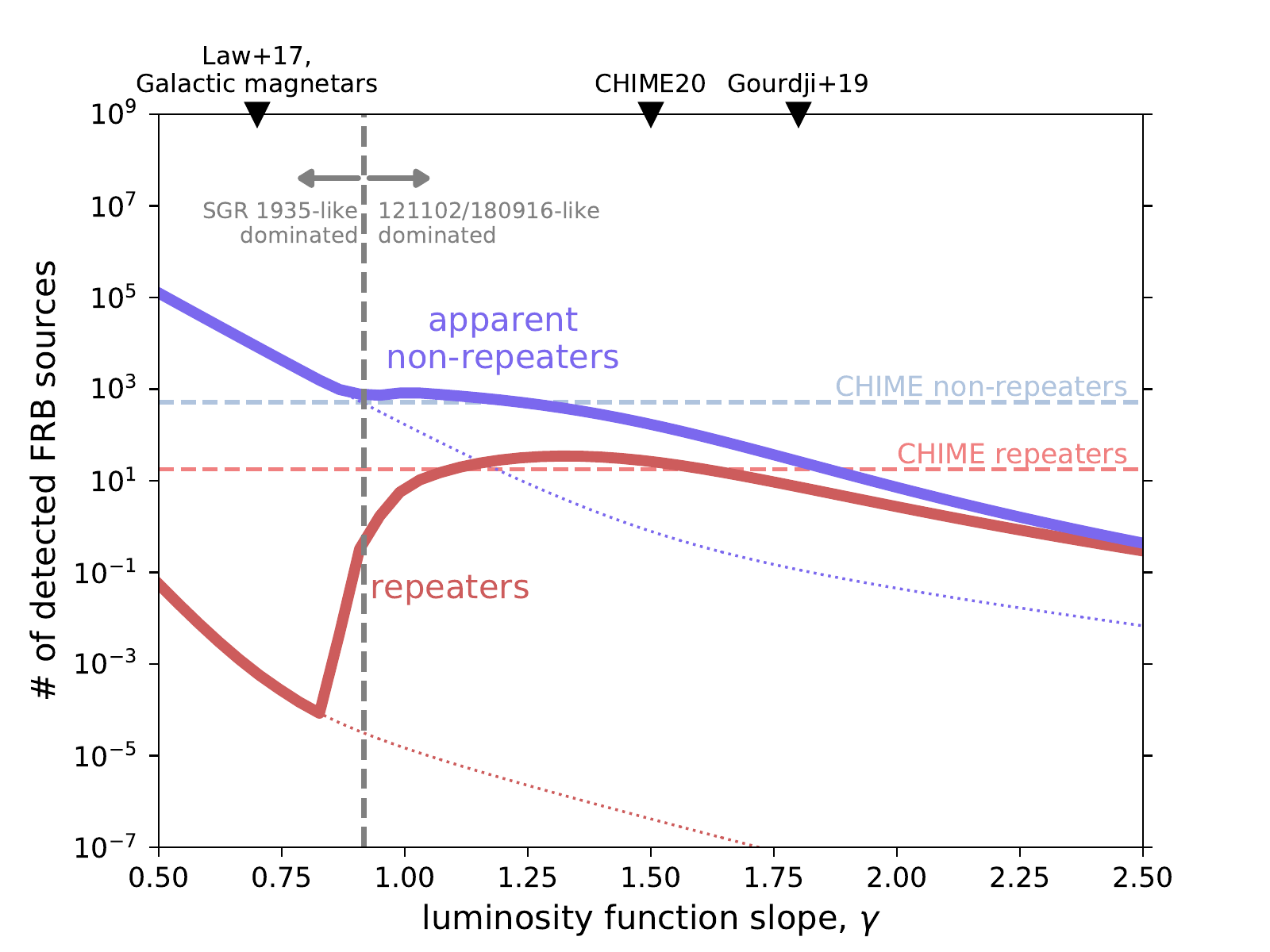}
    \includegraphics[width=0.45\textwidth]{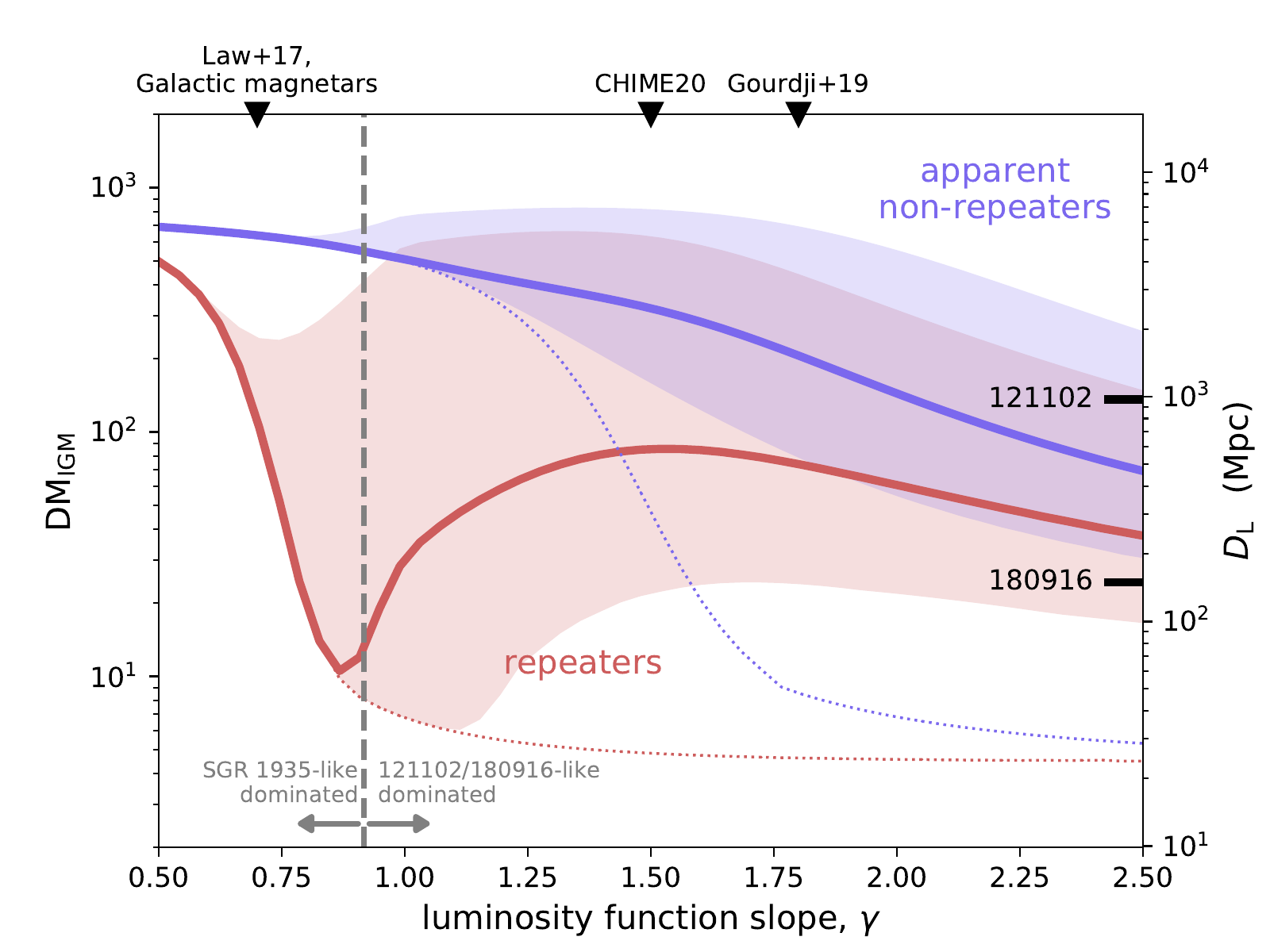}
    \caption{Several properties of cosmological FRBs can be simultaneously reproduced by postulating the existence of a rare population of magnetars (with a birth rate taken to be $f_{\rm CCSN} = 10^{-4}$ of the CCSN rate in this example) with stronger magnetic fields than the Galactic magnetar population. 
    {\it Left:} 
    For a mock survey with parameters motivated by the CHIME/FRB experiment,
    the number of detected repeating sources (solid red) and apparently-non-repeating sources (solid blue), as a function of the power-law slope $\gamma$ of the FRB luminosity function $\lambda(>E) \propto E^{-\gamma}$.  The number of repeating and non-repeating sources actually detected by CHIME are shown as light blue and red dashed horizontal curves, respectively (V.~Kaspi, private communication). The thin dotted curves show the contribution of the ``ordinary magnetar'' population to the number of detected sources. This population is subdominant for $\gamma \gtrsim 1$ and dominant at low $\gamma$, where the all-sky FRB rate is over-produced (Fig.~\ref{fig:FRB_rate}).
    {\it Right:} Median distance (right axis) and corresponding DM contribution from the intergalactic medium (left axis) of sources detected as repeaters (solid red) and non-repeaters (solid blue) for the same mock survey. The distances of FRB 180916 and FRB 121102 (which from CHIME's point of view is a ``non-repeater'') are shown for comparison.
    }
    \label{fig:Bfield}
\end{figure*}

Figure~\ref{fig:Bfield} shows $N_{\rm rep}$ and $N_{\rm non-rep}$ as a function of $\gamma$, and for a representative value of $f_{\rm CCSN}=10^{-4}$ that is consistent with the constraints on FRB~121102-like activity and the all-sky FRB rate (Fig.~\ref{fig:repetition_rate}).
The number of detected repeating and non-repeating sources can be compared to values from the CHIME FRB survey, shown as horizontal curves (V.~Kaspi, private communication). The figure shows that, for values $1 \lesssim \gamma \lesssim 1.5$, the absolute number and relative ratio of repeating and non-repeating FRBs can be reproduced simultaneously with the all-sky rate and per-source activity rate.
At low values of $\gamma$ the observed FRBs are dominated by the less-active ``ordinary magnetar'' population. This results in a significant reduction in the relative number of repeating versus non-repeating sources that is inconsistent with observations.
This substantiates our previous claim that a single population of (or equivalently, a population dominated by) \source-like magnetars cannot account for the number of known repeaters.

Finally, using the cumulative distribution of detected events implied by eqs.~(\ref{eq:Nnonrep},\ref{eq:Nrep}), we calculate the characteristic distances at which repeating and non-repeating FRBs would be detected by this mock survey.
The right panel of Fig.~\ref{fig:Bfield} shows the median distance of detected repeating and apparently-non-repeating FRB sources as a function of $\gamma$.
Confirmed repeaters are detected on-average at a lower distance, broadly consistent with the $149 \, {\rm Mpc}$ distance of the first localized CHIME repeater (and note that FRB~121102, at a much larger distance of $972 \, {\rm Mpc}$ is detected only once by CHIME, consistent with the median distance of apparently-non-repeating sources detected by the mock survey).

As the model shows, many potential rare magnetar formation channels
could in principle be consistent with the all-sky FRB and repeater rate constraints (Fig.~\ref{fig:repetition_rate}).  One way to further break this degeneracy is via host galaxy demographics.  A high-$B$ (and hence potentially particularly slowly rotating; \citealt{Beniamini+20}) tail of the magnetar population should be formed in otherwise ordinary CCSNe and hence track star-forming galaxies almost exclusively.  Superluminous supernovae (SLSNe) and long-duration gamma-ray bursts (LGRB) should originate predominantly \citep{Fruchter+06,Lunnan+15,Blanchard+16}, though not exclusively (e.g.~\citealt{Perley+17}), in dwarf star-forming galaxies.
By comparison, NS mergers, white dwarf-NS mergers, and accretion-induced collapse (AIC) should originate from a
range of star-forming and non-star-forming galaxies \citep{Margalit+19},
weighted more towards massive galaxies than SLSNe/LGRB.  Attempts to
perform an analysis along these lines are already underway
(e.g.~\citealt{Margalit+19}; \citealt{Li&Zhang20}), though it should be cautioned that
without at least arcsecond localization, it is usually challenging to
uniquely identify the host galaxy \citep{Eftekhari+18}, much less the local
environment within the host galaxy as becomes available with VLBI localization \citep{Bassa+17,Marcote+20}.

\section{Implications for Magnetar FRB models}
\label{sec:magnetar_models}

\source~provides a clear link between FRBs and magnetars.  Such a connection has been proposed and discussed extensively in the FRB literature well prior to this discovery \citep[e.g.][]{Popov&Postnov13,Lyubarsky14,Kulkarni+14,Katz16,Metzger+17,Beloborodov17,Kumar+17}, leading to the development of {\it several distinct magnetar models} for FRBs.  Broadly speaking, these models can be further divided based on whether the radio emission originates from near the NS magnetosphere (the ``curvature'', ``low twist'', and ``reconnection'' models) or at much further distances (``synchrotron maser blastwave'' models).  We also briefly discuss a couple NS-related models not specific to magnetars.  In the following, we discuss implications of the \source~radio burst for these models, pointing out strengths and points of contention between each and the combined radio and X-ray observations.

\subsection{Low Twist Models}

In the low twist models \citep{Wadiasingh2019,Wadiasingh2020}, magnetic field dislocations and oscillations in the NS surface can lead to pair cascades assuming that the background charge density is sufficiently low. The latter then result in coherent radio emission. Since it is the same dislocation that supposedly results in short X-ray bursts and (in some cases) FRBs, a prediction of this model is that all FRBs should be associated with short magnetar bursts (but not vice versa). This is consistent with the observations of SGR~1935. Another attractive feature of this model, is that FRBs will typically be associated with older magnetars, consistent with the putative age of SGR~1935. However, the association with older ages in the model is due to the requirement of longer periods, while the period of SGR 1935+2154 is very typical as compared to other Galactic magnetars. Similarly, the model favours strong dipole field strengths, while the dipole field of SGR 1935+2154 does not appear to be exceptional relative to other magnetars.

In the low twist model, both the radio emission and the X-rays arrive from the magnetosphere. As such, this model predicts a radius-to-frequency mapping leading to frequency selection, similar to the case in pulsars and potentially a polarization angle that is varying with time (as the magnetic field orientation changes relative to the observer). The X-ray burst associated with an FRB should exhibit a standard blackbody (or double blackbody) spectrum as seen in other short X-ray bursts.

\subsection{Synchrotron Maser Blastwave Models}

In synchrotron maser models, a version of which was first proposed by \cite{Lyubarsky14}, FRBs are created via the coherent synchrotron maser process that is naturally produced in magnetized relativistic shocks \citep{Gallant+92,Plotnikov&Sironi19}. 
Such shocks are expected to arise from relativistic flares that may be ejected during magnetar outbursts.
A number of variants on the synchrotron maser model exist which differ regarding the nature of the upstream medium and the required shock properties; however, in all cases the bursts are powered by tapping into a small fraction of the kinetic energy of the outflow and predict corresponding (though differing) high-frequency counterparts to FRBs.

\subsubsection{Magnetar Wind Nebula (Lyubarsky 2014)}
\label{sec:MWN}

\citet{Lyubarsky14} propose FRB production occurs as the ultra-relativistic flare ejecta collides with the pulsar wind nebula. The radius of the termination shock is estimated to be
\begin{equation}
    r_{\rm s} = \sqrt{\frac{L_{\rm sd}}{4\pi p c}},
\end{equation}
where $p$ is the pressure inside the nebula.  Taking $L_{\rm sd} \simeq 1.7\times 10^{34}$ erg s$^{-1}$ and $p \sim 10^{-9}$ erg cm$^{-3}$ (estimated from the energy of $10^{51}$ erg of a typical supernova and the observed $\sim 20 \, {\rm pc}$ size of the supernova remnant surrounding \source; \citealt{Kothes+18}), we find $r_{\rm s} \approx 6\times 10^{15}$ cm.  
The light crossing timescale to this radius is $r_{\rm s}/c \sim 2$ day.  However, because the flare ejecta and the resulting shock are also moving close to the speed of light, radio photons from the shocked nebula could in principle still arise nearly simultaneously with the X-rays, which in this model presumably must be generated from the inner magnetosphere.\footnote{Although a burst of higher frequency (incoherent) synchrotron  is predicted in this model from the shocked electrons, for parameters appropriate to \source~this is predicted to occur at $\gtrsim$ GeV gamma-rays.}

Looking more closely at the predictions for the radio emission requires rescaling the results of \citet{Lyubarsky14}, who considered a giant flare which carries away a significant fraction of the magnetic energy of the star.   The recent flare from \source~was less energetic by a factor of $\sim 10^{6}$, corresponding to a strength of the magnetic field in the pulse smaller by a factor of $\sim 10^{3}$, i.e. dimensionless constant $b \sim 10^{-3}$ in the notation of \citet{Lyubarsky14}.  Following equation (8) of \citet{Lyubarsky14} for these parameters, the predicted peak frequency of the maser emission from the forward shock is estimated to be $\nu_{\rm pk} \sim 10$ MHz.  Given the drop-off in the $\nu L_{\nu}$ spectrum of the maser from $\nu_{\rm pk}$ to $\sim 100 \nu_{\rm pk}$ by a factor of $\sim 10^{-4}$ \citep{Plotnikov&Sironi19}, the fraction of the total radio emission in the 1.4~GHz band of STARE2 would be only $\sim 10^{-4}$.  Given an intrinsic efficiency of the synchrotron maser emission of $f_\xi \sim 10^{-2}-10^{-3}$ (see next section), the resulting net efficiency of the radio emission of $\sim 10^{-6}-10^{-7}$ is at best marginally consistent with the observations if the energy of the flare ejecta were comparable to the released X-ray fluence.   

\subsubsection{Baryonic Shell (Metzger et al. 2019, Margalit et al. 2020)}
\label{sec:MMS_model}

In this version of the synchrotron maser model, first proposed by \cite{Beloborodov17}, the ultra-relativistic head of the magnetar flare collides not with the magnetar wind nebula, but instead with matter ejected from a recent, earlier flare.  Motivated by the inference from the radio afterglow of the 2004 giant flare from SGR 1806-20 of a slow ejecta shell generated by the burst \citep{Gelfand+05,Granot+06}, \citet{Metzger+19} consider the upstream medium to be a sub-relativistic baryon-loaded shell with an electron-ion composition.  

The low efficiency of radio emission implied the X-ray and radio observations of \source~ (eq.~\ref{eq:energy_ratio}) is consistent with predictions of this model.  The radio inefficiency is attributable to a combination of the intrinsic synchrotron maser efficiency $f_\xi \sim 10^{-2}-10^{-3}$ (for moderate magnetization; \citealt{Plotnikov&Sironi19}) and a further reduction by a factor of $\sim 10^{-2}$ due to the effects of induced-Compton scattering suppressing the low-frequency portion of the maser's intrinsic SED (see \citealt{Metzger+19}, their $\S$3.2).\footnote{The value of $f_{\xi}$ in general decreases with increasing values of the upstream magnetization, $\sigma$.  Based on 1D particle-in-cell simulations of electron-positron plasmas, \citet{Plotnikov&Sironi19} find an efficiency of
\begin{equation}
f_{\xi} = 7\times 10^{-4}/\sigma^{2}, \sigma \gg 1.
\end{equation}
Matching $f_{\xi} = \eta \sim 10^{-5}$ (eq.~\ref{eq:energy_ratio}) places a strict upper limit $\sigma \lesssim 8$.  The true efficiency (and hence allowed $\sigma$) will be lower once accounting for 3D effects, and electron-ion composition of the upstream plasma (e.g.~\citealt{Iwamoto+19}), and suppression of the radio signal from induced scattering by upstream electrons \citep{Metzger+19}.  This scenario therefore requires an upstream plasma which is not highly magnetized.}

Following the methodology of \citet{Margalit+20} we can use the energy, frequency, and duration  of the observed radio burst to derive the intrinsic parameters of the flare demanded by the synchrotron maser model.  Adopting the observed quantities from $\S\ref{sec:observations}$, we find that the energy of the relativistic flare $E_{\rm flare}$; the Lorentz factor $\Gamma$ of the shocked gas at the time the observed radio flux is emitted; the radius of the shock from the central magnetar at this time, $r_{\rm sh}$; and the external density of the upstream baryonic shell at this location, $n_{\rm ext}$, are given by
\begin{eqnarray}
    &&E_{\rm flare} \approx 10^{40} \, {\rm erg} \, f_{\xi,-3}^{-4/5}  \left(\frac{W}{5 \, {\rm ms}}\right)^{1/5} d_{10}^2,
    \\
    &&\Gamma \approx 24 \, f_{\xi,-3}^{-1/15} \left(\frac{W}{5 \, {\rm ms}}\right)^{-2/5} d_{10}^{1/3},
    \\
    &&n_{\rm ext} \approx 9.3 \times 10^4 \, {\rm cm}^{-3} \, f_{\xi,-3}^{-4/15}  \left(\frac{W}{5 \, {\rm ms}}\right)^{2/5} d_{10}^{-2/3},
    \\
    &&r_{\rm sh} \approx 1.7 \times 10^{11} \, {\rm cm} \, f_{\xi,-3}^{-2/15}  \left(\frac{W}{5 \, {\rm ms}}\right)^{1/5} d_{10}^{2/3}.
\label{eq:shockparams}
\end{eqnarray}
In the above we express results in terms of the $1.4 \, {\rm GHz}$ burst duration, $W$, which we normalize to $5 \, {\rm ms}$ motivated by the quoted CHIME burst width \citep{Scholz_ATel13681}, in addition to the maser efficiency $f_{\xi} = 10^{-3}f_{\xi,-3}$. 

From our inferred parameters, the (very) local DM contributed by the immediate upstream medium ahead of the shock is ${\rm DM} \gtrsim n_{\rm ext} r_{\rm sh} \approx 5 \times 10^{-3} \, {\rm pc \, cm}^{-3}$, and can exceed this value significantly if the upstream medium extends to larger radii. In the context of the shock model, it may be expected that DM variations could exist between radio bursts on this order of magnitude or larger. Note however, that non-linear wave interaction with the upstream plasma as well as strong upstream magnetic fields may inhibit the DM of this local environment \citep[e.g.][]{LuPhinney19}.

\begin{figure}
    \centering
    \includegraphics[width=0.45\textwidth]{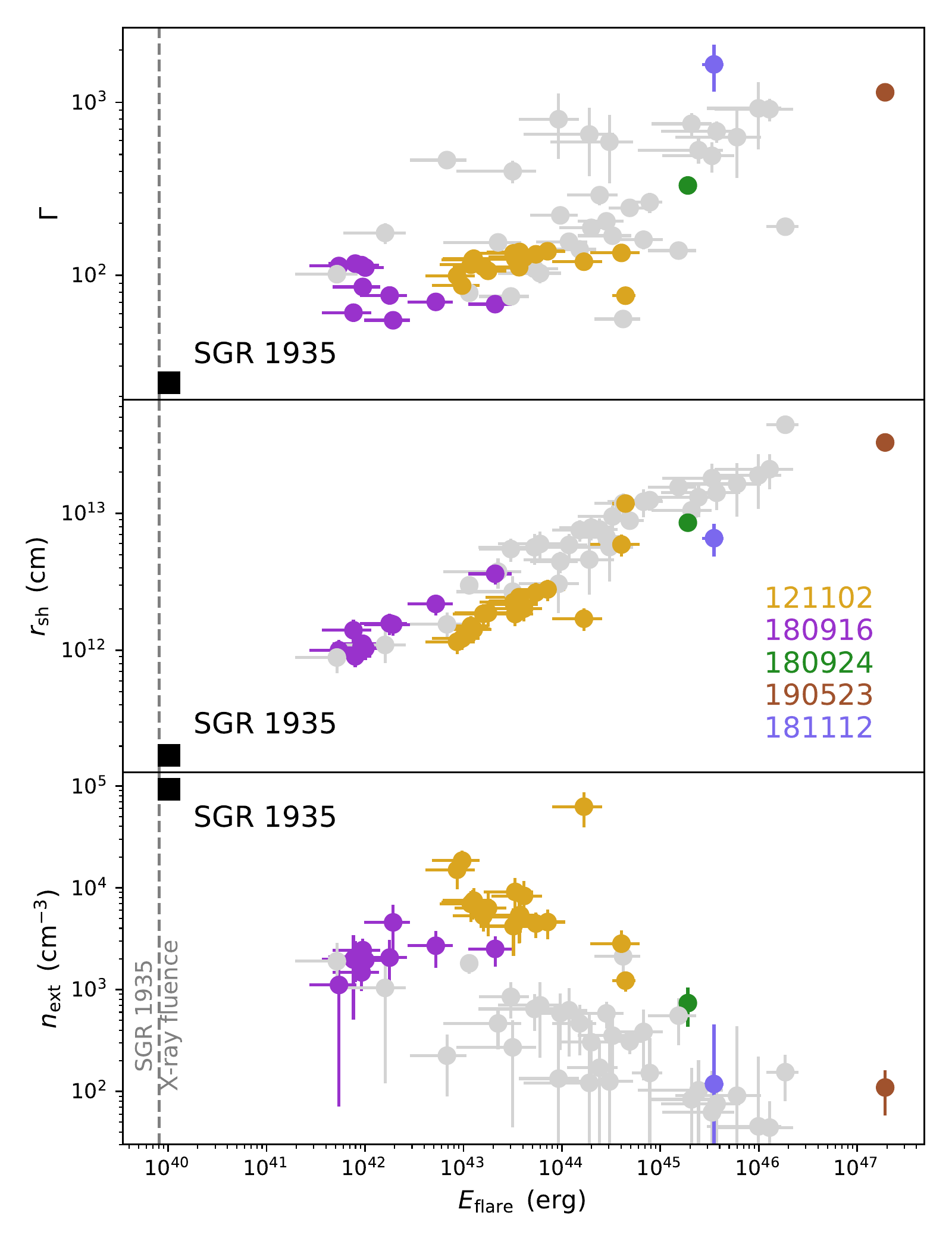}
    \caption{Derived properties from the SGR~1935 radio burst within the variation of the synchrotron maser model in which the upstream medium is a baryon-loaded shell and adopting a fiducial value for the maser efficiency $f_{\xi} = 10^{-3}$ \citep[see also][]{Margalit+20}.  From top to bottom, the bulk Lorentz factor $\Gamma$, radius $r_{\rm sh}$, and the external upstream density $n_{\rm ext}$ of the shock at the time of the observed radio flare, all as a function of the derived total flare energy, $E_{\rm flare}$.  The flare energy, as derived from radio observations alone, is comparable to the detected X-ray counterpart of this burst (vertical dashed curve), in-line with predictions of the synchrotron maser model (\S\ref{sec:MMS_model}) if the X-rays arise from thermal synchrotron radiation from the same shocks which generate the FRB.  
    }
    \label{fig:shock_derived_properties}
\end{figure}

The derived shock properties are shown in Fig.~\ref{fig:shock_derived_properties} in comparison to those derived for cosmological FRBs within the same model, for an assumed efficiency $f_{\xi} = 10^{-3}$.   One important thing to note is that the flare energy $E_{\rm flare}$ the model demands (based on the radio observation of \source~alone) agrees remarkably well with the independently observed X-ray energy, $E_{\rm X} \approx 8 \times 10^{39} \, {\rm erg}$.  
As we discuss below, such agreement is naturally expected if electrons heated at the shock generate the X-rays via synchrotron radiation in the fast-cooling regime.  

Stated more directly, the model predicts a ratio \citep{Margalit+19} 
\begin{eqnarray}
\eta_{\rm theory} &\equiv& \frac{E_{\rm radio}}{E_{\rm flare}} = \left(\frac{1215}{512\pi^{2}}\frac{m_{\rm e}}{m_{\rm p}}\right)^{1/5}f_{\rm e}^{1/5}f_{\xi}^{4/5}\left(\nu_{\rm obs}\cdot t_{\rm FRB}\right)^{-1/5} \nonumber \\
&\sim& 3\times 10^{-5}\left(\frac{f_{\rm e}}{0.5}\right)^{1/5}f_{\xi,-3}^{4/5}\left(\frac{\nu_{\rm obs}\cdot t_{\rm FRB}}{1{\rm GHz\cdot ms}}\right)^{-1/5},
\label{eq:eta_theory}
\end{eqnarray}
which matches the observed ratio $\eta \equiv E_{\rm radio}/E_{\rm X} \sim 10^{-5}$ (eq.~\ref{eq:energy_ratio}) for expected values of the maser efficiency $f_{\xi} \sim 10^{-3}-10^{-2}$ \citep{Plotnikov&Sironi19} provided that $E_{\rm X} \sim E_{\rm flare}$. 
In the above, $f_{\rm e}$ is the ratio of electron to ion number densities in the upstream medium, and is $f_{\rm e} \sim 1$ for an electron-ion plasma as we consider.

The same outwardly propagating shock that generates the coherent precursor radio emission (the ``FRB") in this scenario also generates incoherent synchrotron radiation from relativistically-hot electrons heated by the shock \citep{Lyubarsky14,Metzger+19}, somewhat akin to a gamma-ray burst afterglow.  Using the shock parameters implied by the radio observations (Fig.~\ref{fig:shock_derived_properties}), we now assess the predicted properties of the high-frequency counterpart, showing it to be in accord with the X-rays emission from~\source.

The peak frequency of the ``afterglow" is set by the characteristic synchrotron frequency, which for an ultra-relativistic blastwave decelerating into an effectively stationary upstream medium of magnetization $\sigma = 10^{-1} \sigma_{-1}$ is given by (\citealt{Metzger+19}; their eqs.~56, 57)
\begin{eqnarray}
    E_{\rm peak} &\sim& h \nu_{\rm syn} \approx 50\,{\rm keV}\left(\frac{E_{\rm flare}}{10^{39}{\rm erg}}\right)^{1/2}\sigma_{-1}^{1/2}\left(\frac{t}{{\rm 5 \,ms}}\right)^{-3/2}  \nonumber \\
    &\approx& 160 \, {\rm keV} \, \sigma_{-1}^{1/2} \left(\frac{W}{5 \, {\rm ms}}\right)^{1/10} \left(\frac{t}{5 \, {\rm ms}}\right)^{-3/2} d_{10},
\label{eq:Epeak}
\end{eqnarray}
where $t$ is the time since the peak of the flare and we have assumed that $1/2$ of the kinetic power dissipated at the shock goes into heating electrons into a relativistic Maxwellian distribution (supported e.g. by particle-in-cell simulations of magnetized shocks; \citealt{Sironi&Spitkovsky11}).  In the second line of equation (\ref{eq:Epeak}) we have substituted the flare energy from equation (\ref{eq:shockparams}) needed to reproduce the radio burst properties from \source.
Although the value of $\sigma$ in the flare ejecta of a magnetar flare is uncertain theoretically, its value is nevertheless reasonably constrained: a minimum magnetization $\sigma \gtrsim 10^{-3}$ is required for the synchrotron maser to operate in the first place, while the declining efficiency of the maser emission with increasing $\sigma$ places an upper limit $\sigma \lesssim 1$.  

For the same parameters, the cooling frequency is given by
\begin{eqnarray}
    h \nu_{\rm c} 
    &\approx& 13 \, {\rm keV} \, \sigma_{-1}^{-3/2} \left(\frac{n_{\rm ext}}{10^{4}\rm cm^{-3}}\right)^{-3/2} \left(\frac{\Gamma}{100}\right)^{-4} \left(\frac{t}{\rm ms}\right)^{-2}
    \nonumber \\
    &\approx& 5.6 \, {\rm keV} \, \sigma_{-1}^{-3/2} \left(\frac{W}{5 \, {\rm ms}}\right)^{-1} \left(\frac{t}{5 \, {\rm ms}}\right)^{-1/2} d_{10}^{-1/3},
\end{eqnarray}
where the particular temporal scaling $\propto t^{-1/2}$ is derived assuming a radially-constant density profile ($n_{\rm ext} \propto r^0$).  

The fact that $\nu_{\rm c} \lesssim \nu_{\rm syn}$ on timescales $t \sim $ 5 ms of interest shows that the post-shock electrons are fast-cooling and hence a large fraction of the flare energy, $E_{\rm flare}$, is emitted as hard X-rays of energy $E_{\rm peak} \sim 10-100$~keV. The predicted X-ray spectrum is thus fast-cooling sychrotron emission from relativistically-hot electrons with a thermal Maxwellian energy distribution (see \citealt{Giannios&Spitkovsky09}, their Fig.~3), resulting in an ordinary fast-cooling spectrum $\nu L_{\nu} \propto \nu^{1/2}$ between $\nu_{\rm c}$ and $\nu_{\rm syn}$ and an exponential cut-off at an energy $\sim \nu_{\rm syn}$.  Indeed, modeling of bright magnetar flares, suggests that they can be well fit by a cut-off power-law spectrum in the X-rays \citep{vdH2012}.  

\begin{figure}
    \centering
    \includegraphics[width=0.45\textwidth]{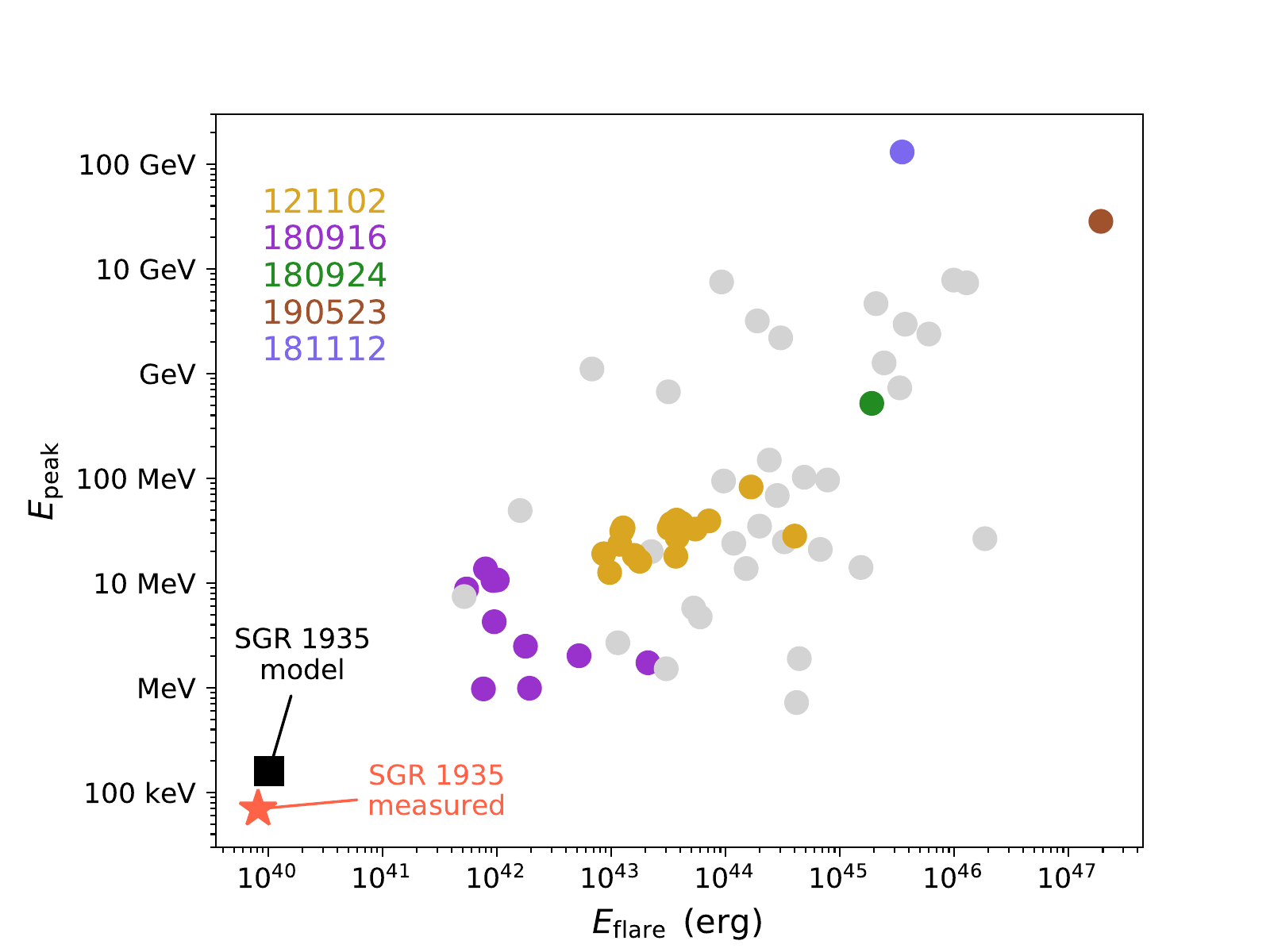}
    \caption{The synchrotron maser model predicts that FRBs be accompanied by hard radiation counterparts, the properties of which can be derived by modelling the observed radio emission alone. Above, we show the peak energy of this counterpart as a function of flare energy, derived for a sample of cosmological FRBs and the \source~radio burst, assuming a uniform magnetization $\sigma = 0.1$ for the upstream medium.  The emitting electrons are fast cooling and therefore $E_{\rm flare}$ roughly corresponds to the radiated fluence at $E_{\rm peak}$. The observed $E_{\rm peak}$ and fluence of the contemporaneous X-ray burst associated with \source's radio emission agree remarkably well with the model prediction. }
    \label{fig:Epeak}
\end{figure}

Extending the same model to the shock properties derived for the observed populations of cosmological FRBs predicts that the afterglow emission for these more energetic bursts will occur at much higher energies $E_{\rm peak} \gtrsim$ MeV-GeV in the gamma-ray band (Fig.~\ref{fig:Epeak}).  Unfortunately, gamma-ray satellites like {\it Swift} and {\it Fermi} are generally not sensitive enough to detect this emission to the cosmological distances of most FRB sources \citep{Metzger+19,Margalit+20,Chen+20}. We furthermore emphasize that this predicted short ($\sim$milliseconds duration) gamma-ray signal from the shocks is distinct from the longer-lasting and typically softer gamma-ray emission observed from giant Galactic magnetar flares (e.g.~\citealt{Palmer+05,Hurley+05}), which is instead well explained as a pair fireball generated by dissipation very close to the NS surface.  The latter being relatively isotropic compared to the relativistically-beamed radio emission from the ultra-relativistic shocks (hypothesized to accompany the beginning of the flare; \citealt{Lyubarsky14,Beloborodov17}) might explain the non-detection of FRB-like emission from the 2004 giant flare of SGR 1806-20 \citep{Tendulkar+16}. 

If the X-rays from magnetar flares are attributable to thermal synchrotron shock emission, this may be imprinted in correlations between X-ray observables.  Since electrons behind the shock are fast-cooling, the X-ray fluence $F_{\rm X}$ should scale linearly with the flare energy $E_{\rm flare}$, from which one predicts from equation (\ref{eq:Epeak}) a correlation (for fixed $\sigma$) 
\begin{equation}
E_{\rm peak} \propto F_{\rm X}^{1/2}t_{\rm X}^{-3/2},
\label{eq:correlation}
\end{equation}
between the spectral energy peak, burst fluence and some measure of the burst duration $t_{\rm X}$.  

For X-ray bursts from the Galactic magnetar SGR J1550-5418, \citet{vdH2012} report a correlation between the fluence and ``emission" time, $\tau_{\rm 90} \propto F_{\rm X}^{0.47}$.  Taking $t_{\rm X} \propto \tau_{90}$ then equation (\ref{eq:correlation}) predicts $E_{\rm peak} \propto F_{X}^{-0.2}$, which is close to but slightly shallower than the correlation $E_{\rm peak} \propto F_{X}^{-0.44}$ found by \citet{vdH2012} using the entire burst sample.  However, note that the most energetic bursts studied by \citet{vdH2012}  exhibit a flatter or even positive correlation of $E_{\rm peak}$ with fluence.   A correlation very close to $E_{\rm peak} \propto F_{X}^{-0.44}$ is also predicted from equation (\ref{eq:correlation}) using the empirical relationship $F_{\rm X} \propto t_{\rm X}^{1.54}$ found between the bursts from different magnetars (\citealt{Gavriil+04}; see Fig.~\ref{fig:burst_duration}).  Thus, we advance the radical hypothesis that hard X-ray emission in even ordinary magnetar flares can be generated by internal shocks in baryon-loaded outflows.  

Scaling the shock model to the lower values of $r_{\rm sh},\Gamma$ than derived for~\source~above, would also have potential consequences for the X-ray spectrum.  In particular, the synchrotron spectrum could in some X-ray counterparts of Galactic magnetar flares become opaque to pair creation.  Following \cite{Lithwick2001,BG2017} we calculate the pair creation opacity corresponding to the model parameters described above, and find the external shock to be optically thin to pair creation. Pair creation could however become important for somewhat lower values of $r_{\rm sh},\Gamma$, which could be realized if the dissipation is due to internal shocks at a radius that is much lower than $r_{\rm sh}$. For example, keeping $E_{\rm flare}$ fixed and reducing the Lorentz factor to $\Gamma \sim 3$, will lead to a pair-creation cutoff at the internal shock site at $\sim 40$ keV.  We speculate that this might explain why the ``high temperature" of X-ray bursts described by a double black-body spectrum typically peak at energies $\lesssim40$keV. 

Finally, the explanation provided here for the X-ray spectra of Galactic magnetar short bursts might apply most robustly only to the smaller sub-set of bright SGR bursts which are luminous enough to drive radiation-driven outflows (i.e. large values of $L_{\rm X}/L_{\rm min}$, see Fig.\ref{fig:burst_epoch}).  In this regard it is interesting that the brightest bursts analyzed by \cite{vdH2012} exhibit a different correlation between their peak energy and flux to the rest of the bursts, consistent with the idea that the physical mechanism driving these bursts is different.

Returning to the radio burst emission, \citet{Metzger+19} predict a downwards drift in frequency\footnote{This phenomena, which is now well cataloged for many repeating FRB sources, has been termed the ``sad trombone'' \citep{Hessels+19,CHIME_repeaters}.} as the shock decelerates and the (Lorentz boosted) synchrotron maser emission sweeps from high to low frequencies.  The presence (or lack of) this feature in the \source~burst would therefore provide a helpful diagnostic of FRB models and a probe of the density profile in this specific burst (see \citealt{Margalit+20} for application to CHIME repeaters).  In addition to the in-band drift that may potentially be detectable by CHIME (although the non-trivial frequency response of the CHIME sidelobes may hinder this), the same process could manifest as a small arrival time-delay between the STARE2 detection at $1.4 \, {\rm GHz}$ and the CHIME detection at lower frequencies.

The relative delay between the observed radio emission (corrected for DM) and it's hard-radiation counterpart observed at peak (Fig.~\ref{fig:Epeak}) should be, at most, comparable to the radio burst duration ($\lesssim 5 \, {\rm ms}$ in the case of \source's April 28$^{\rm th}$ burst). This results from the fact that the high-frequency photons are optically thin at the shock deceleration radius (when the emission peaks), while the radio is optically-thick to induced scattering at this time, and only escapes at a time $t \sim$burst-width later.

\begin{figure}
    \centering
    \includegraphics[width=0.45\textwidth]{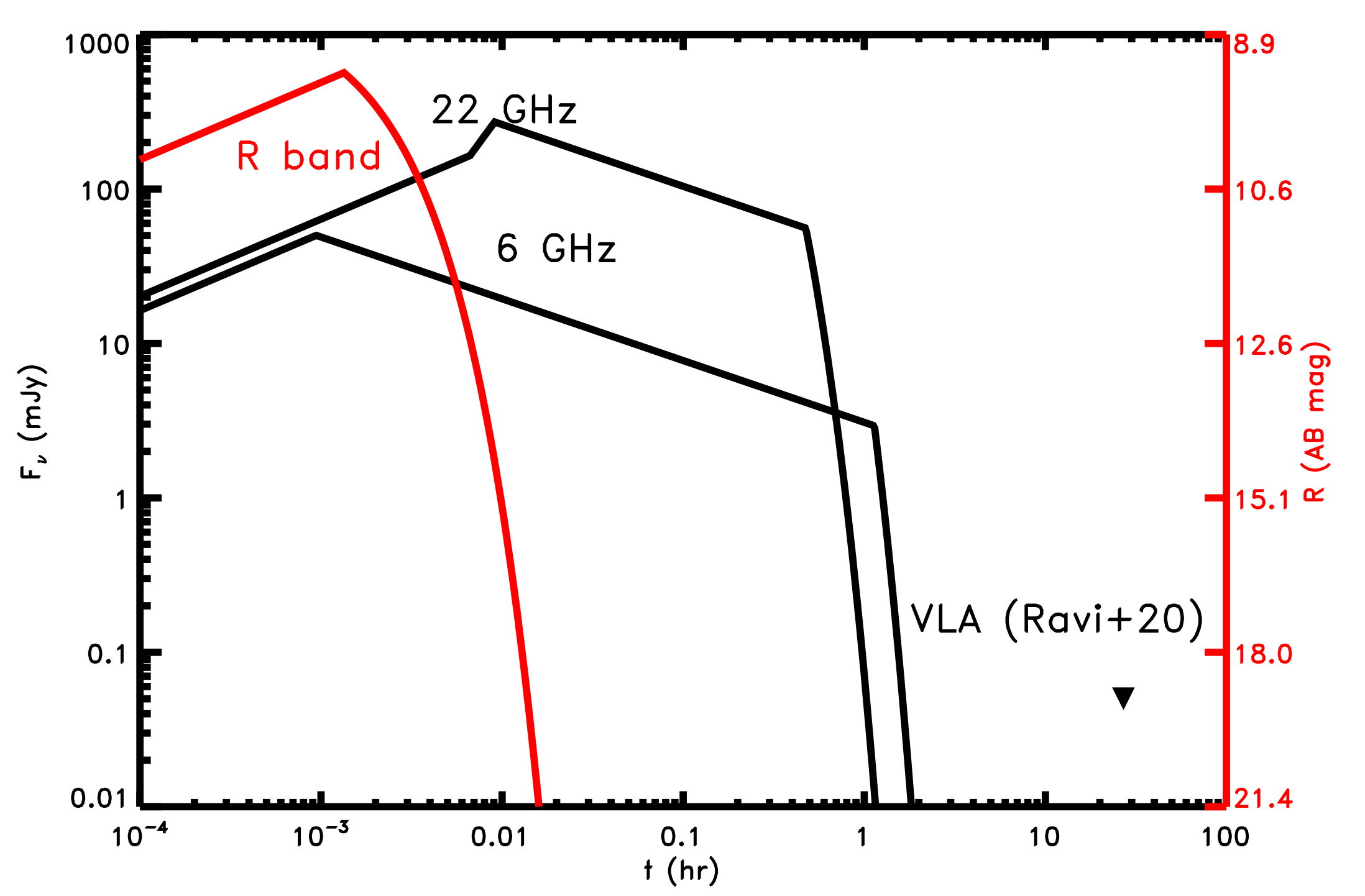}
    \caption{Optical R-band and 6/22 GHz radio flux due to thermal synchrotron afterglow emission of the X-ray/FRB-generated shock as a function of time in hours (bottom axis) as it propagates to larger radii.  Shown for comparison are upper limits from \citet{Ravi_ATel13690}.  In this simple estimate we have assumed a radially-constant density of the external medium; if the external medium does not extend so far then the predicted flux would be lower than these estimates. Synchrotron self-absorption, which affects the radio light curves, has been included in an approximate way.}
    \label{fig:afterglow}
\end{figure}

As shown in Figure \ref{fig:afterglow}, the shock model predicts that the X-ray emission should be accompanied by longer-lived synchrotron emission at lower frequencies as the shock propagates to larger radii, again similar to a GRB afterglow \citep{Sari+98}.  For frequencies below $\nu_{\rm syn}$ and $\nu_{\rm c}$, the peak flux will rise as $F_{\nu} \propto \nu^{1/2}$ as $\nu_{\rm syn} \propto t^{-\alpha}$ decreases in time until reaching the observing frequency, where $\alpha = 3/2$ during the relativistic phase and $\alpha = 3$ once the blastwave transitions to become non-relativistic.  Thus, if $\nu_{\rm syn}$ is passing through the X-ray band ($\nu_{\rm X} \sim 10$ keV) on a timescale of $t_{\rm X} \sim 0.1$ s, it will reach the optical band on a timescale of seconds and the radio band ($\nu_{\rm R} \sim 10$ GHz) on a timescale of less than an hour.  For an external medium with a radially constant density, the flux density $F_{\nu}$ at peak ($\nu_{\rm obs} \sim \nu_{\rm syn}$) is constant during the relativistic phase and thus approximately equal to that achieved initially at the higher frequencies.  Using as a proxy for the latter the X-ray fluence of $F_X \sim 7\times 10^{-7}$ erg/cm$^{2}$ of the FRB-generating flare from \source, we predict a peak radio afterglow flux density $F_{\rm radio} \sim F_{X}/(\nu_{\rm X}\cdot t_{\rm X}) \sim 0.3$~Jy (although self-absorption severley mitigates this estimate; see Fig.~\ref{fig:afterglow}).  After peak, the predicted flux will decay exponentially as the synchrotron emission occurs from electrons on the Wien tail of the thermal Maxwellian.  Thus, such an afterglow is consistent with upper limits on radio afterglow emission of 0.05 mJy taken on a timescale of a few days \citep{Ravi_ATel13690,Ravi_ATel13693}.  Also note that the radius sampled by the shock at these late times is a factor $\gtrsim 30$ times larger than at the time of the earlier radio emission and there is no guarantee that the dense medium extends that far from the magnetar. 

Finally, the ejection of the baryonic outflow is predicted to result in a small decrease in the spin frequency of the magnetar due to the temporary opening up of field lines by the mass-loaded wind \citep{BT1998,Harding1999,Beniamini+20}. The magnitude of the spin frequency change increases with the luminosity of the baryonic wind and its duration, both of which are highly uncertain.  However if the X-ray burst concurrent with the FRB produced such an outflow, then there are at least two other bursts within the latest outburst, with even larger luminosities \citep{Veres_GCN27531,Ridnaia_GCN27631} that may have also produced an outflow. Taking for example the brightest of these bursts, that occurred on April 22$^{\rm nd}$, we can calculate a lower limit on the spin frequency decrease of $\Delta \Omega=2\times 10^{-8} L_{41}^{1/2}t^{1/2}\mbox{ s}^{-1}$ \citep{Beniamini+20}.

\subsubsection{Spin-Down Powered Wind (Beloborodov 2017, 2019)}
\label{sec:Beloborodov}

\citet{Beloborodov17, Beloborodov19} argue that the upstream medium into which the relativistic flare collides is an electron/positron plasma from a spin-down powered component to the pulsar wind.  
Given the low spin-down power of \source~however, this scenario might be somewhat disfavored. 

The observed radio fluence at $\nu = 1.4 \, {\rm GHz}$ in this model can be estimated from equation 91 of \cite{Beloborodov19}, and re-expressed as a function of the spin-down power $L_{\rm sd}$ and pair multiplicity $\mathcal{M}$,
\begin{equation}
\label{eq:Beloborodov_Eradio}
    E_{\rm radio} \sim 3 \times 10^{31} \, {\rm erg} \,  L_{{\rm sd},34}^{1/12} \mathcal{M}_3^{2/3} .
\end{equation}
In the above we have assumed the flare energy is $10^{40} \, {\rm erg}$ motivated by the X-ray counterpart of \source, and we have omitted scaling with nuisance parameters for clarity.
For the spin-down power of \source~and standard pair-multiplicity of $10^3$, this fluence is $\sim$three orders of magnitude lower than the observed energy of \source's radio burst.

This tension may be alleviated by an enhancement of the magnetar-wind (because of e.g. opening of field lines by the magnetar flare) or an increased pair-multiplicity shortly preceding the flare \citep{Beloborodov19}.
From equation~(\ref{eq:Beloborodov_Eradio}) we find that a pair-multiplicity of $\sim 5 \times 10^7$ would be required to fit the observed fluence. This is much higher than pulsar pair multiplicities, though it has been suggested that much larger values $\gg 10^3$ may be attainable for magnetars immediately preceding a flare \citep{Beloborodov19}.

The high pair-loading in this version of the synchrotron maser model, as in the \cite{Lyubarsky14} model, push the high-energy thermal synchrotron counterpart to lower frequencies ($\nu_{\rm syn} \propto f_{\rm e}^{-2}$), leading to an ``afterglow" peaking optical/UV band \citep{Beloborodov19} instead of the x-ray/gamma-ray band predicted in the baryonic model (Fig.~\ref{fig:Epeak}).  In this scenario, as in the magnetar wind nebula case ($\S\ref{sec:MWN}$), the X-rays from the flare must be created by a process which is not directly related to the FRB  mechanism.

\subsection{Additional Models}
\label{sec:additional_models}

In curvature models, the FRB is produced by curvature radiation from bunched electrons streaming along the magnetic field lines of the magnetar \citep[e.g.][]{Kumar+17,Lu&Kumar18}. These models predict fluence ratios $\eta \sim 1$ in all bands \citep{Chen+20}, and thus cannot explain properties of the observed X-ray burst of \source~(whose fluence ratio is $\eta \sim 10^{-5} \ll 1$, but also in terms of X-ray spectrum) as a bona fide FRB counterpart.
This shortcoming can potentially be dismissed by arguing that the same magnetar activity leading to particle bunching, acceleration, and FRB emission, also produces ``normal'' short X-ray bursts. However, this scenario makes no prediction as to the quantitative relationship between the radio and X-ray burst properties, and the observed fluence ratio of $\eta \sim 10^{-5}$ is ad-hoc within this framework.
These considerations apply also to models where FRBs are produced by reconnection in the outer magnetosphere \citep{Lyubarsky20}, or indeed to any magnetospheric models.
If true, a testable prediction of the scenario (and one that would be in tension with synchrotron-maser models) may be large variations in the value of $\eta$ for future events.


Another class of FRB models discussed in the literature power FRBs by the kinetic energy of an outflow interacting with the magnetosphere of a NS \citep{Zhang17,Zhang18,IokaZhang20}. The outflow in these ``cosmic comb'' models has been suggested to range from AGN-driven winds, to GRB jets, or winds from a binary companion to the NS. None of the above are likely to be applicable in the context of \source's radio burst, in tension with predictions of this model \citep[although see][]{Wang+20}.

A final class of NS-related models we discuss are ``spin-down models'', in which the radio bursts are powered by rotational energy of the NS \citep{CordesWasserman16,Connor+16,Munoz+20}. Though in principle applicable to magnetars, these models typically envision normal pulsars as progenitors. Indeed, considering the very low spin-down power of \source~in comparison to pulsars, and the fact that the latter are $\gtrsim 100$ times more common than Galactic magnetars (in terms of their effective active lifetime), it seems unnatural that a spin-down-powered ``FRB'' would be first detected for \source.
A prediction of this model would be a change in NS spin period due to the released burst energy.
Accounting for the radiated energy of \source's burst, including it's associated X-ray counterpart, we find that a change $\Delta \Omega \approx -4 \times 10^{-6} \, {\rm s}^{-1} d_{10}^2$ in the angular velocity of \source~need to have occurred.
A reduction in the spin frequency is also expected in the \cite{Metzger+19} version of the synchrotron maser model due to opening of magnetic field lines by the baryonic outflow, however this effect is  much smaller (\S \ref{sec:MMS_model}). The \cite{Beloborodov19} synchrotron maser model (\S\ref{sec:Beloborodov}) also necessitates some level of period change due to a significant enhancement of pulsar-wind required immediately preceding the FRB, however this too would be expected to be smaller than $\Delta \Omega$ estimated above.
Future timing of \source~may help test this prediction of the spin-down model. 

\section{Conclusions}
\label{sec:conclusions}

We have explored implications related to the population of FRB sources and to theoretical models of FRBs in light of the recent detection of a luminous millisecond radio burst from the Galactic magnetar \source~\citep{Scholz_ATel13681,Bochenek_ATel13684}.
The large energy of this burst makes it unique amongst any previously observed pulsar/magnetar phenomenology, and bridges the gap to extragalactic FRBs \citep{Bochenek_ATel13684}.

Our conclusions may be summarized as follows.  With regards to general implications for extragalactic magnetar populations:
\begin{itemize}
\item Broadly speaking, the discovery of a highly luminous millisecond-duration radio burst coincident with the X-ray flare of the Galactic magnetar \source~supports magnetar models for extragalactic FRBs.  

\item The X-ray properties of the flare are fairly typical of Galactic magnetar flares in terms of fluence and overall duration (Fig.~\ref{fig:burst_duration}), however it's X-ray spectrum may have been harder than usual \citep{Mereghetti+20}.  Furthermore, with the notable exception of the giant flare from SGR 1806-20, relatively few such flares would have previously been detected at radio wavelengths for a similar low ratio $\eta \equiv E_{\rm radio}/E_{\rm X} \sim 10^{-5}$ of radio to X-ray fluence (Fig.~\ref{fig:burst_epoch}).  This suggests that a sizable fraction of Galactic magnetar flares may be accompanied by luminous radio bursts, although we caution that multiple emission mechanisms may be at play in producing magnetar X-ray flares, and that only the subset produced by shocks would be expected to emit coherent radio bursts. 

\item Applying the same fluence ratio $\eta$ to giant magnetar flares would imply that Galactic magnetars are capable of powering even the most energetic cosmological FRBs (Fig.~\ref{fig:luminosity_function}).  However, a stark discrepancy exists between the activity (burst repetition rate) of Galactic magnetars and the sources of the recurring extragalactic FRBs.  If universal, the low efficiency $\eta \sim 10^{-5}$ also places strong upper limits on the magnetar active lifetime in the latter case, much shorter than the ages of Galactic magnetars (eq.~\ref{eq:tactive}).  

\item The estimated rate of radio bursts similar to that observed from \source~is insufficient to contribute appreciably to the observed extragalactic FRB rate (eq.~\ref{eq:rate_basic}).  Depending on the luminosity function of the Galactic flares, the all-sky FRB rate (including also giant flares) can be reproduced by ``ordinary'' Galactic magnetars similar to \source~(Fig.~\ref{fig:FRB_rate}).  However, such a model fails to simultaneously explain the large (per-source) repetition rate of known repeaters or the large DMs of the FRB population.

\item Instead, considering a two-component model---we add a second population of magnetars whose birth-rate is a free-parameter, and whose activity level and lifetime are  scaled from \source~as a function of the population's magnetic field strength. 
This model allows to broadly replicate the observed properties of the FRB population (Fig.~\ref{fig:Bfield}), however only if the birth-rate of the ``active'' magnetar population is $\ll$ than the CCSN rate (Fig.~\ref{fig:repetition_rate}; in line with previous work, e.g. \citealt{Nicholl+17}).

\item This implies that the population of ``active'' magnetars cannot be interpreted as an earlier evolutionary state of \source-like magnetars which are born in a large fraction of CCSNe. Instead this population may form through more exotic channels such as SLSNe, AIC, or binary NS mergers \citep{Metzger+17,Margalit+19}.

\end{itemize}

In addition to the general implications summarized above, we also address implications for specific FRB magnetar models. We stress that there is no single ``magnetar model'' for FRBs, and that many distinct models have been suggested in the literature, and these differ in the requisite magnetar properties, the FRB emission mechanism, and predictions for (of lack of) multi-wavelength counterparts (\S\ref{sec:magnetar_models}). In this context, we find that:
\begin{itemize}

\item Magnetospheric models (``curvature'', ``low twist'', and ``reconnection'' models) predict either no high-energy counterpart, or weak counterparts of comparable energy to the radio emission ($\eta \sim 1$). This is in tension with the observed X-ray counterpart in \source, where $\eta \sim 10^{-5}$. 
This shortcoming can be dismissed by interpreting the X-ray counterpart as a ``normal'' short X-ray burst. However this scenario makes no predictions of the spectral or energetic properties of the X-ray flare and the $\eta \sim 10^{-5}$ radio-to-X-ray fluence ratio is ad-hoc within this framework (\S\ref{sec:additional_models}).
    
\item Synchrotron maser models, which involve relativistic flare ejecta colliding with an external medium, provide a promising alternative.  However, these models differ in the nature of the upstream medium and their predicted multi-wavelength afterglow.
    
\item Models in which the upstream is the magnetar wind nebula (\citealt{Lyubarsky14}; \S\ref{sec:MWN}) may have an efficiency issue and predict associated afterglow in the GeV range.  Likewise, models in which the upstream is a rotational-powered pulsar wind \citep{Beloborodov17,Beloborodov19} are strained and predict a lower frequency counterpart (\S\ref{sec:Beloborodov}).
    
\item The baryonic shell version of the synchrotron maser model naturally explains the value of $\eta$ (compare eqs.~\ref{eq:energy_ratio} and \ref{eq:eta_theory}) in addition to the timing\footnote{Note that the X-rays and radio are predicted to arrive contemporaneously in this model, contrary to arguments made by \citet{Mereghetti+20}, who did not account for relativistic time-of-flight effects.} and spectral features of the observed X-ray emission (Fig.~\ref{fig:Epeak}). The model requires substantial mass ejection to accompany the flares, which may be in tension with the relatively low luminosity of the flares if the outflows are driven by radiation pressure (Fig.~\ref{fig:burst_epoch}).  On the other hand, the requirement for mass ejection---and the sensitivity of the radio emission to the detailed properties of the upstream medium---could help explain why not all magnetar flares are accompanied by a luminous FRB. 
    
\item The latter model suggests a new paradigm in which X-ray emission of magnetar flares arises from thermal synchrotron radiation from internal shocks.  This model predicts a power-law spectrum with an exponential cut-off, and correlations between the peak energy of the burst and other burst properties (e.g. duration and fluence), which are broadly consistent with observations \citep{vdH2012}.

\item The baryon shell synchrotron maser shock model \citep{Metzger+19,Margalit+20} makes several predictions testable by future Galactic or extragalactic FRBs: (1) Although X-ray/gamma-ray emission can arise from magnetar flares without an accompanying FRB (e.g. if the X-rays/gamma-rays do not arise from shocks), the opposite is not true.  Any FRB-like burst should be accompanied by X-ray/gamma-ray emission with an energy at least a factor $\eta^{-1} \gtrsim 10^{4}$ larger than the emitted radio energy (eq.~\ref{eq:eta_theory}); (2) scaling up to cosmological FRBs, the equivalent prompt synchrotron counterpart should peak in the $\gtrsim$ MeV-GeV gamma-rays band (Fig.~\ref{fig:Epeak}); (3) Galactic FRBs may be accompanied by longer lived optical/radio emission on a timescale of seconds/minutes (Fig.~\ref{fig:afterglow}), the details of which however depend on the extent of the external medium surrounding the magnetar on larger radial scales than probed by the hard X-rays.
    
\end{itemize}

\acknowledgements
We thank Bryan Gaensler and George Younes for helpful insight.
BM is supported by NASA through the NASA Hubble Fellowship grant \#HST-HF2-51412.001-A awarded by the Space Telescope Science Institute, which is operated by the Association of Universities for Research in Astronomy, Inc., for NASA, under contract NAS5-26555. The research of PB was funded by the Gordon and Betty Moore Foundation through Grant GBMF5076. NS is supported by the Columbia University Dean's fellowship and through the National Science Foundation (grant\# 80NSSC18K1104). BDM is supported in part by the Simons Foundation through the Simons Fellowship program in Mathematics and Physics (grant \#606260).

\newpage    
\bibliography{bib}{}
\bibliographystyle{aasjournal}



\end{document}